\documentclass[numberedappendix,twocolappendix,iop,apj]{emulateapj}
\usepackage{psfig,amsfonts,amsmath,graphicx,natbib,nicefrac,longtable,graphicx
}

\usepackage[backref,breaklinks,colorlinks,citecolor=blue,urlcolor=blue]{hyperref}
\usepackage{breakurl}

\usepackage{booktabs}

\newcommand{\tjet}{\ensuremath{t_{\rm jet}}}
\newcommand{\thetajet}{\ensuremath{\theta_{\rm jet}}}

\newcommand{\nua}{\ensuremath{\nu_{\rm a}}}

\newcommand{\numax}{\ensuremath{\nu_{\rm m}}}

\shorttitle{Radio Linear Polarization of GRB Afterglows I}
\shortauthors{Laskar et al.}
\submitted{Accepted for publication in The Astrophysical Journal}

\def\bath{1}
\def\naoj{2}
\def\jao{3}
\def\nrao{4}
\def\naojf{*}

\begin{document} 

\title{Radio linear polarization of GRB afterglows: Instrumental Systematics in ALMA observations of GRB 171205A}
\author{Tanmoy~Laskar\altaffilmark{\bath}}
\author{Charles L.~H.~Hull\altaffilmark{\naoj,\jao,\naojf}}
\author{Paulo Cortes\altaffilmark{\nrao,\jao}}

\affil{\altaffilmark{\bath}Department of Physics, University of Bath, Claverton Down, Bath, BA2 
7AY, United Kingdom}
\affil{\altaffilmark{\naoj}National Astronomical Observatory of Japan, NAOJ Chile, Alonso de 
C\'ordova 3788, Office 61B, 7630422, Vitacura, Santiago, Chile}
\affil{\altaffilmark{\jao}Joint ALMA Observatory, Alonso de C\'ordova 3107, Vitacura, Santiago, 
Chile}
\affil{\altaffilmark{\nrao}National Radio Astronomy Observatory, Charlottesville, VA 22903, USA}
\affil{\altaffilmark{\naojf}NAOJ Fellow}

\begin{abstract}
Polarization measurements of gamma-ray burst (GRB) afterglows are a promising means of probing the 
structure, geometry, and magnetic composition of relativistic GRB jets. However, a 
precise treatment of instrumental calibration is vital for a robust physical interpretation of 
polarization data, requiring tests of and validations against potential instrumental systematics. 
We illustrate this with ALMA Band 3 (97.5~GHz) observations of GRB~171205A taken $\approx5.19$~days 
after the burst, where a detection of linear polarization was recently claimed.
We describe a series of tests for evaluating the stability of polarization measurements with 
ALMA. 
Using these tests to re-analyze and evaluate the archival ALMA data, 
we uncover systematics in the polarization calibration at the $\approx0.09\%$ level. We derive a 
3$\sigma$ upper limit on the linearly polarized intensity of $P<97.2~\mu$Jy, corresponding to an 
upper limit on the linear fractional polarization of $\Pi_{\rm L}<0.30\%$, in contrast to the 
previously claimed detection. Our upper limit improves upon existing constraints on the intrinsic 
polarization of GRB radio afterglows by a factor of 3.
We discuss this measurement in the context of constraints on the jet magnetic field geometry. We 
present a compilation of polarization observations of GRB radio afterglows, and demonstrate that a 
significant improvement in sensitivity is desirable for eventually detecting signals polarized at 
the $\approx0.1\%$ level from typical radio afterglows.
\end{abstract}


\section{Introduction}
Polarization studies of long-duration GRB afterglows are expected to probe the presence of ordered 
magnetic fields in their jetted outflows as well as the viewing geometry 
\citep{gra03,gk03,rlsg04,gt05,kob17}, yielding crucial constraints on the jet launching mechanism 
and the central engine \citep{lyu09,bt16}. Whereas polarization studies in the optical have revealed 
evidence for structured magnetic fields in the outflow \citep{sms+09,ccb+11,mkca+13,wct+14},
similar studies at radio/millimeter (mm) frequencies have been more limited due to instrumental 
sensitivity constraints \citep{tfk+98,fyb+03,tfbk04,gt05,vdhpdb+14,cg16}. 

The advent of the Atacama Large Millimeter/Sub-millimeter Array (ALMA) is changing the landscape, 
and has resulted in the first detection of polarized emission from GRBs in the radio/mm band, which 
has provided preliminary constraints on the magnetic field structure in GRB jets \citep{lag+19}.
Additionally, \cite{uth+19} claimed a detection of $(0.27\pm0.04)\%$ linear polarization in the 
radio afterglow of GRB~171205A, measured $\approx5.19$~days after the burst with ALMA at 97.5~GHz. 
By assuming an intrinsic polarization of $\approx1\%$, and by ascribing the difference between 
the intrinsic and observed polarization to depolarization by a population of non-accelerated 
electrons, they inferred an acceleration fraction of $f_{\rm acc}\approx0.1$. 

As polarization capabilities with ALMA continue to evolve since the initial commissioning effort 
\citep{nnp+16}, consistent analysis frameworks need to be deployed to interpret polarization 
observations, especially in the case of detections near the threshold of the current instrumental 
systematics. Here, we discuss 
strategies for testing data for these systematics in polarization measurements of faint sources. 
We re-analyze the observations reported in \cite{uth+19}, and demonstrate that the data suffer 
from unremovable, systematic calibration uncertainties.

We report our derived upper limit on the 
polarization of GRB~171205A in Section~\ref{text:almapolcal}. We discuss the implications of the 
upper limit on the magnetic field structure, and compare with previous observations of polarized 
emission for GRB radio afterglows in Section~\ref{text:results}.

\begin{figure*}
 \centering
 \begin{tabular}{cc}
  \includegraphics[width=\columnwidth]{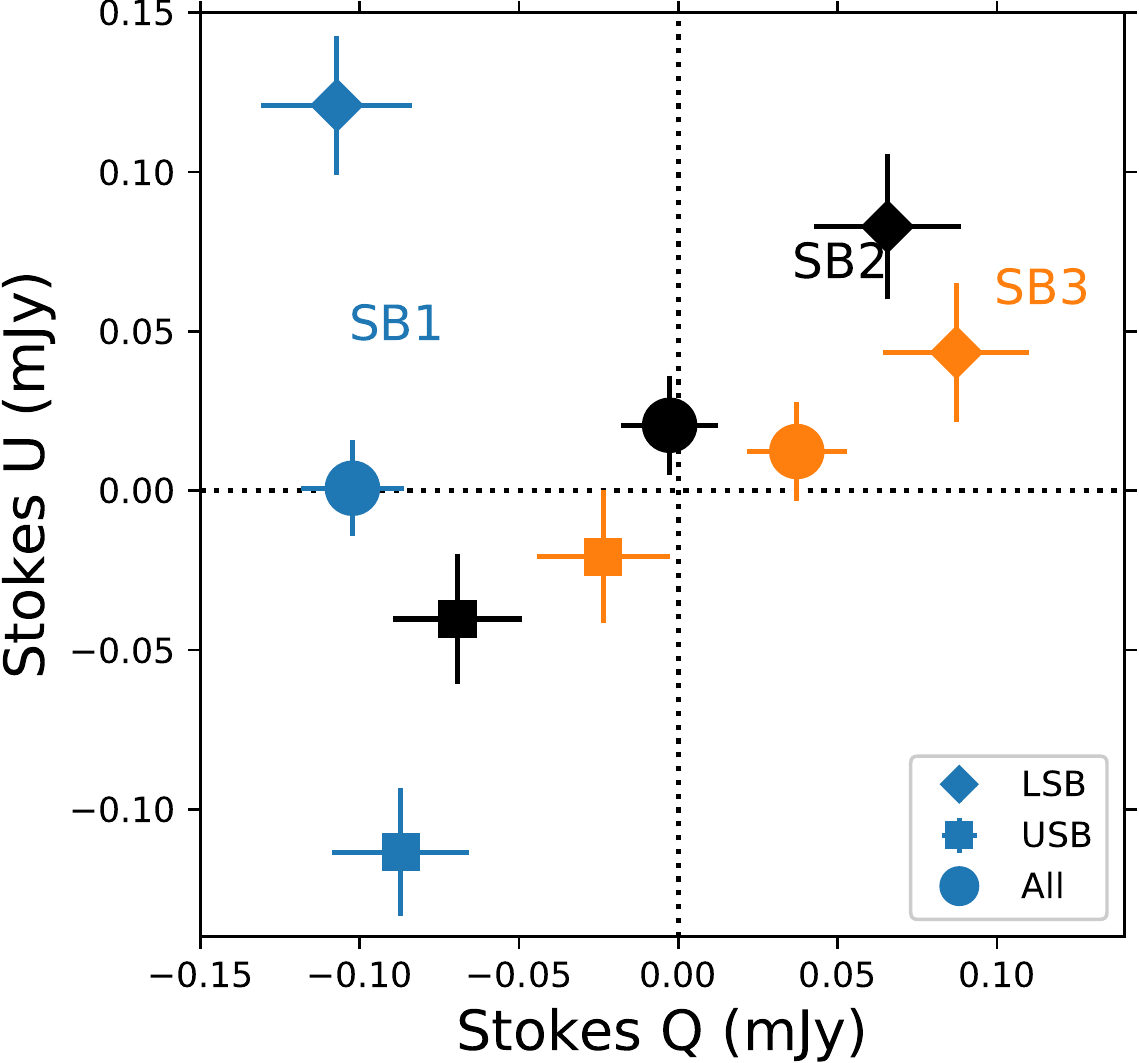} &
  \includegraphics[width=\columnwidth]{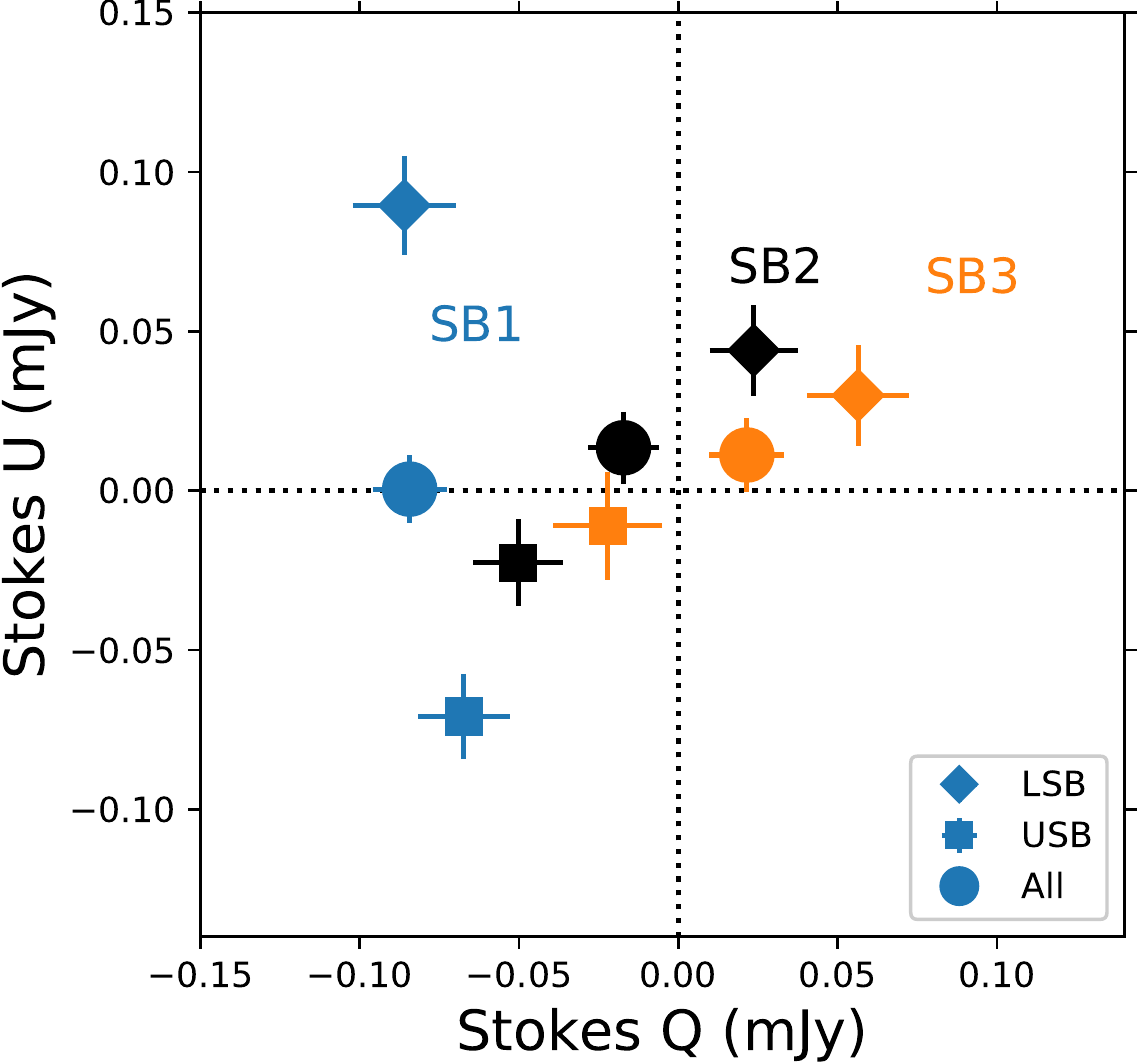} \\
 \end{tabular}
 \caption{Stokes $Q$ versus Stokes $U$ before (left) and after (right) self-calibration for 
GRB~171205A (circles) divided into lower sideband (diamonds) and upper sideband (squares), and 
further by time into the three executions of the scheduling block at 5.136--5.168~d (blue), 
5.174--5.207~d (black), and 5.217--5.245~d (orange). The polarization properties are neither 
consistent in frequency across ALMA Band 3 (points with the same color), nor stable with time 
(points with same marker shape). The $QU$ axis scales are equal and are identical between the two 
panels. Upon self-calibration (see Section~\ref{text:reduction}), 
the uncertainty on the individual measurements is reduced and the 
points shift closer to the origin (zero polarization). The measurements, which span 2.6~hours, 
exhibit an unexpectedly strong trend in time, corresponding to a rotation in the plane of 
polarization from $\approx-44^\circ$ to $\approx31^\circ$ for the self-calibrated data. }
\label{fig:QU_GRB}
\end{figure*}

\section{ALMA polarization observations}
\label{text:almapolcal}
\subsection{QA2 calibration}
We downloaded the 
raw data for full-Stokes ALMA Band 3 (3mm) observations of GRB~171205A taken on 
2017 December 10 under project 2017.1.00801.T (PI: Urata) from the ALMA archive.
The observations employed J1127-1857 as bandpass and flux density calibrator, 
J1256-0547 as polarization calibrator, and J1130-1449 as complex gain calibrator. 
As a first step, we used the CASA \citep{mws+07} calibration scripts, associated with 
the data set and also available from the ALMA archive, to 
regenerate the calibrated Quality Assurance 2 (QA2) measurement set. 
We made images in Stokes $IQUV$ from the full calibrated measurement set with CASA version 5.6.1 
using a robust parameter of 0.0, 
and also independently from the lower sideband (LSB; $89.5$--$93.5$~GHz) and upper 
sideband (USB; $101.5$--$105.5$~GHz) data. The rms noise near the center of the 
$Q$, $U$, and $V$ images is ${\approx7.3}~\mu$Jy, consistent with the expected thermal noise 
given the 
observation duration. The GRB afterglow is well detected in Stokes $I$, with a flux density of 
$30.97\pm0.09$~mJy measured using CASA \texttt{imfit}\footnote{{The uncertainties reported 
by \texttt{imfit} follow the prescription of \cite{con97}.}}. The Stokes $I$ image is dynamic range 
limited, 
with an rms $\approx80~\mu$Jy\footnote{The expected theoretical rms for the full 3-hour observation 
is $\approx7~\mu$Jy.}.
We also detect a point source in maps of Stokes $Q$ and $U$. Fitting for the linearly polarized 
flux density with the position fixed to that derived from the Stokes $I$ image, we obtain 
$Q=-68.8\pm7.3~\mu$Jy and $U=-45.6\pm7.5~\mu$Jy, in agreement with the values reported by 
\cite{uth+19}. However, we find that the Stokes $Q$ measurements differ between the two sidebands 
by 
$32~\mu$Jy, corresponding to a difference in linear polarization fraction, $\Pi_{\rm 
L}\approx0.1\%$ 
relative to Stokes $I$. 

We tested for stability of polarization calibration by dividing the data in time by each execution 
of 
the scheduling block (SB), as described in \cite{lag+19}. This approach reveals systematic trends 
in the $QU$ time evolution. Stokes $Q$ appears to increase from $-87.8\pm12.2~\mu$Jy to 
$-48.2\pm12.2~\mu$Jy (a change of $\approx0.11\%$ of Stokes $I$) over the course of the 
observations, while Stokes $U$ appears to increase from $-82.4\pm12.5~\mu$Jy to $-20\pm13.2~\mu$Jy
($\approx0.19\%$ of Stokes $I$), where the uncertainties refer to those associated with the point 
source fits with \texttt{imfit}, {and which are compatible with the expected thermal noise 
in each SB execution of $\approx12~\mu$Jy.} 
This variability is especially strong in the USB, with $Q$ and $U$ apparently changing by 
$\approx0.23\%$ and $\approx0.30\%$ of Stokes $I$, respectively, over the course of the 
observations (Fig.~\ref{fig:QU_GRB}). 
The magnitude of these temporal changes are much larger than the absolute value of 
the polarization detection previously claimed by \cite{uth+19} with these data. 

We also note the presence of significant signal in circular polarization, with Stokes 
$V=-69.8\pm7.4~\mu$Jy ($\approx0.23\%$ of $I$), at the same level as the previously claimed linear 
polarization detection. Circular polarization has only been reported once in a GRB afterglow 
\citep{wct+14}, and its detection here is more likely indicative of instrumental systematics than 
of an intrinsic origin. 
We note that the observed Stokes $V$ is within the current systematic uncertainty for on-axis 
circular polarization with ALMA ($\approx0.6$\%). 

Finally, we also image the gain calibrator (J1130-1449), dividing the data in time into three bins 
by scheduling block executions. 
The linear polarization properties of the gain calibrator appear to vary over 
the course of the observations, with Stokes $Q$ increasing from $8.87\pm0.04$~mJy (1.11\% of Stokes 
$I$) to $9.79\pm0.04$~mJy (1.23\%; a $\approx3\sigma$ change, corresponding to 0.12\% of $I$) and 
Stokes $U$ increasing from $-14.85\pm0.05$~mJy (1.86\% of $I$) to $-13.12\pm0.06$~mJy (1.64\%; a 
$\approx34\sigma$ change, corresponding to 0.22\% of $I$). The gain calibrator also appears to 
exhibit a statistically significant circular polarization signal, with $V=-1.89\pm0.05$~mJy 
(0.24\% of $I$). These calibrators are not expected to be significantly circularly polarized in the 
mm band, and thus the Stokes $V$ measurement most likely indicates residual polarization 
calibration errors. We discuss this further {in Section \ref{text:polmeasurements}}. 

\begin{deluxetable}{clrrrr}
 \tabletypesize{\footnotesize}
 \tablecolumns{11}
  \tablecaption{Derived polarization properties of the polarization calibrator, J1256-0547}
  \tablehead{   
   \colhead{Reference} &
   \colhead{Method} &
   \colhead{$Q$} &
   \colhead{$U$} &
   \colhead{$\Pi_{\rm L}$\tablenotemark{a}} &
   \colhead{$\chi$\tablenotemark{a}} \\
   \colhead{Antenna} &
   \colhead{}      &
   \colhead{(\%)} &
   \colhead{(\%)} &
   \colhead{(\%)} &
   \colhead{(deg)} 
   }
 \startdata 
 DV06 & qufromgain & $2.08\pm0.06$ & $5.96\pm0.04$ & $6.31$ & $35.4$ \\
      & XYf+QU     & $2.19\pm0.12$ & $5.89\pm0.03$ & $6.29$ & $34.8$ \\
      & residual   &$-0.01\pm0.09$ & $0.07\pm0.08$ & $0.07$ & $47.8$ \\
 DA64 & qufromgain & $2.17\pm0.15$ & $5.99\pm0.09$ & $6.37$ & $35.0$ \\
      & XYf+QU     & $2.20\pm0.12$ & $5.94\pm0.06$ & $6.33$ & $34.8$ \\
      & residual   &$-0.05\pm0.26$ & $0.10\pm0.23$ & $0.11$ & $59.3$ 
 \enddata
 \label{tab:polcal}
 \tablenotetext{a}{$\Pi_{\rm L}$ is the linear polarization fraction and $\chi=\arctan{(U/Q)}$ is 
the polarization (electric field vector) position angle. \texttt{qufromgain} and \texttt{xyamb} do 
not provide uncertainties on these 
quantities}
\end{deluxetable}

\begin{figure*}
  \centering
 \begin{tabular}{ccc}
   \includegraphics[width=0.31\textwidth]{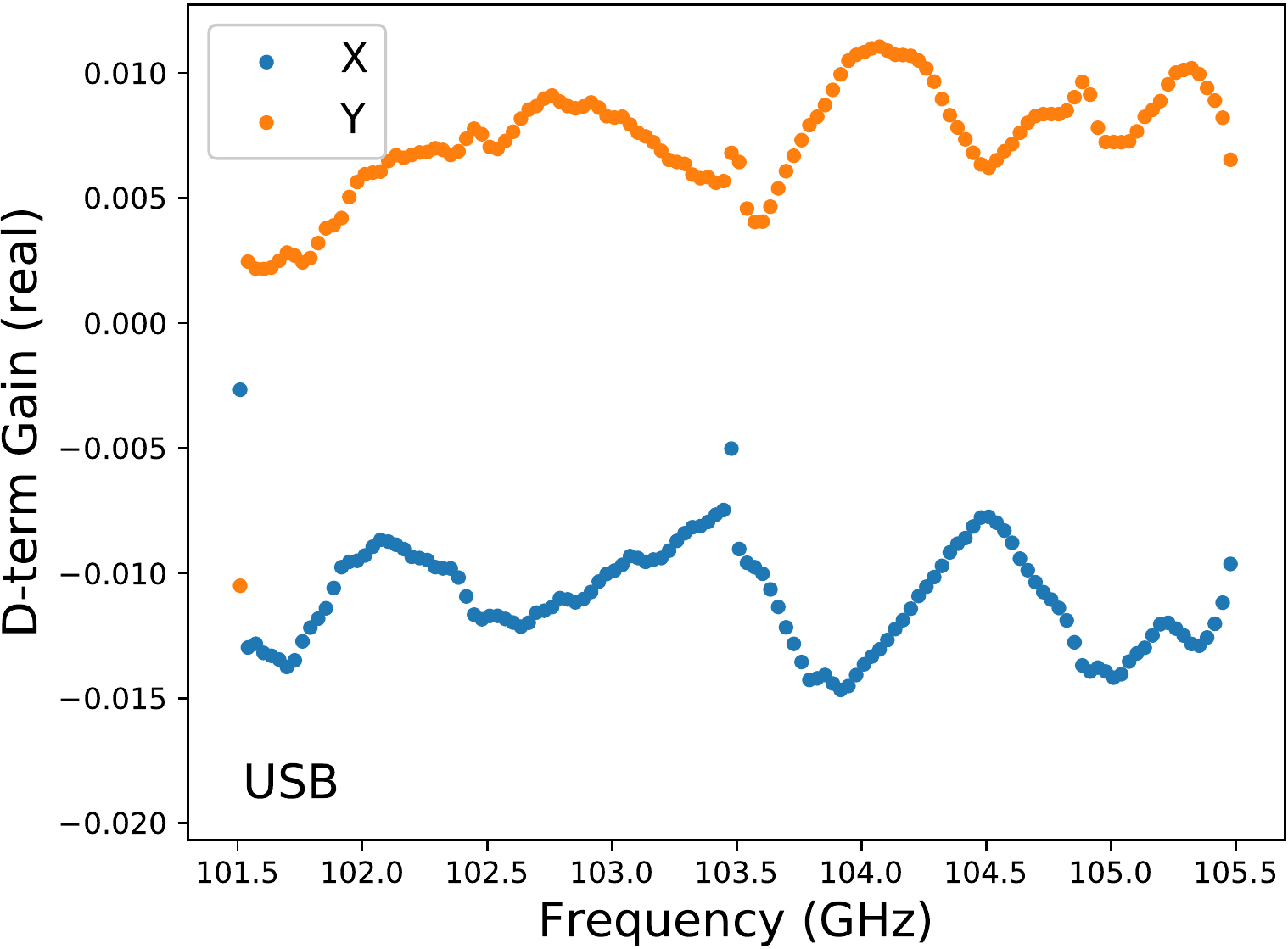}  &
   \includegraphics[width=0.31\textwidth]{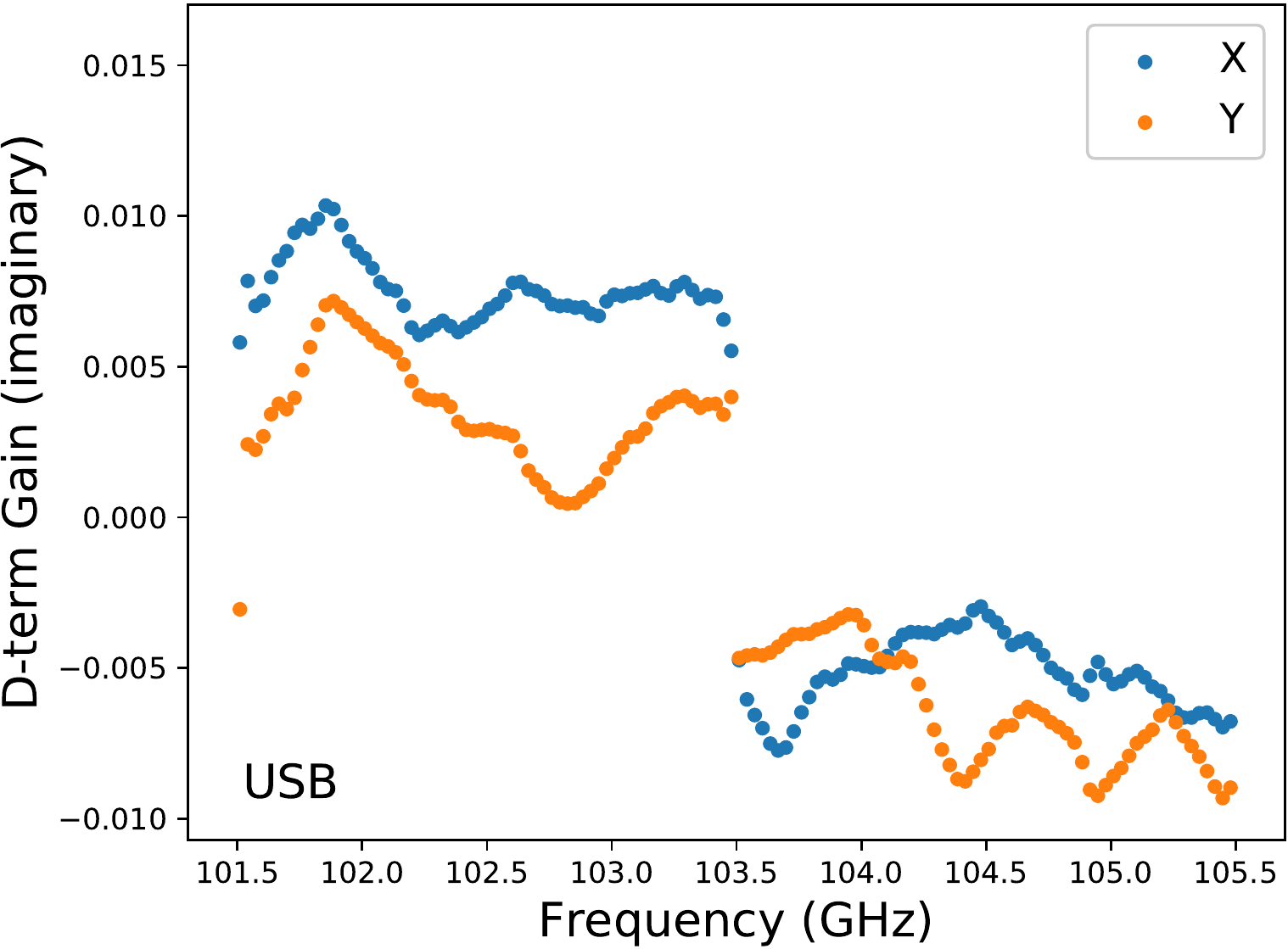}  &
   \includegraphics[width=0.31\textwidth]{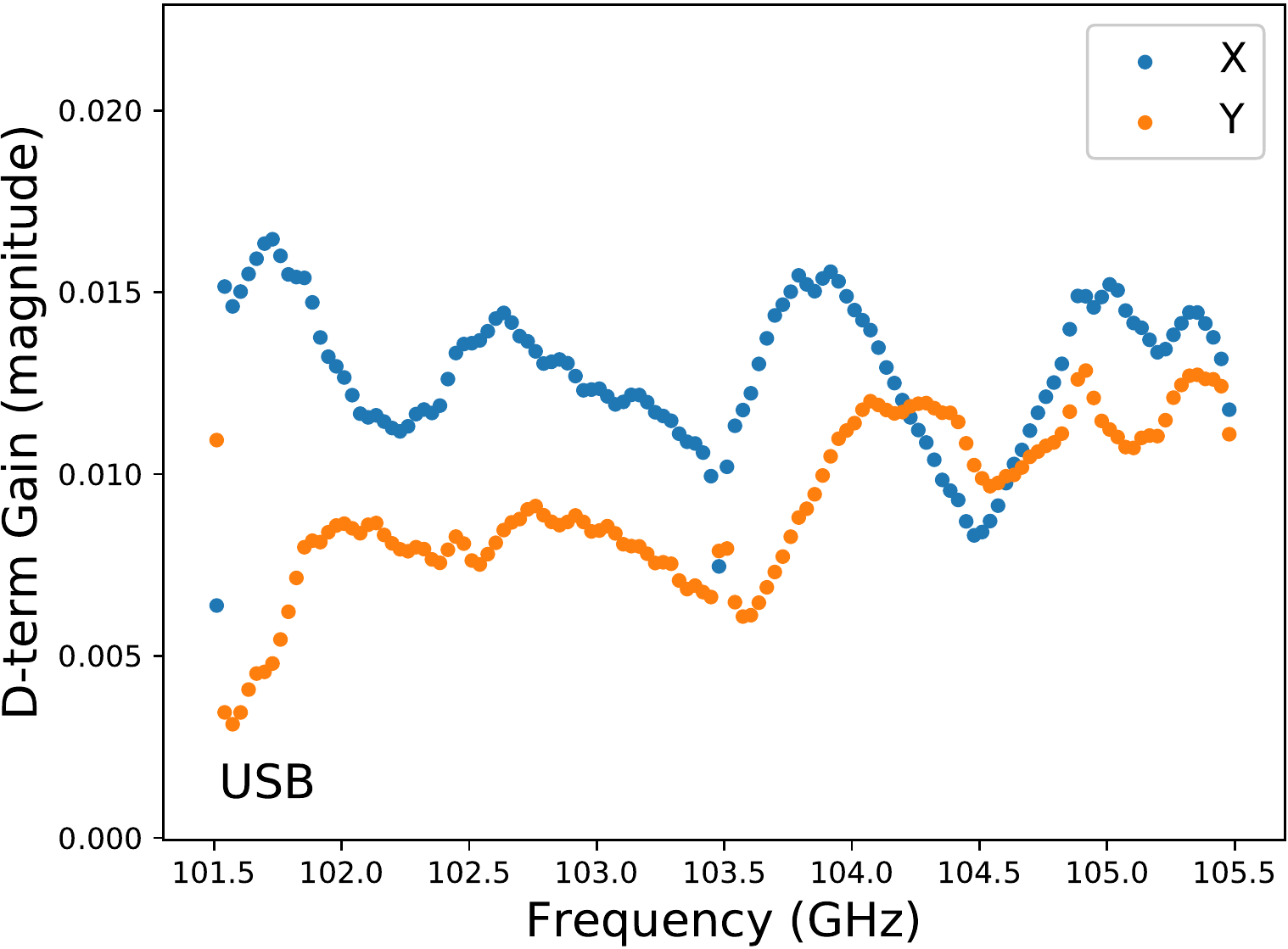} \\
   \includegraphics[width=0.31\textwidth]{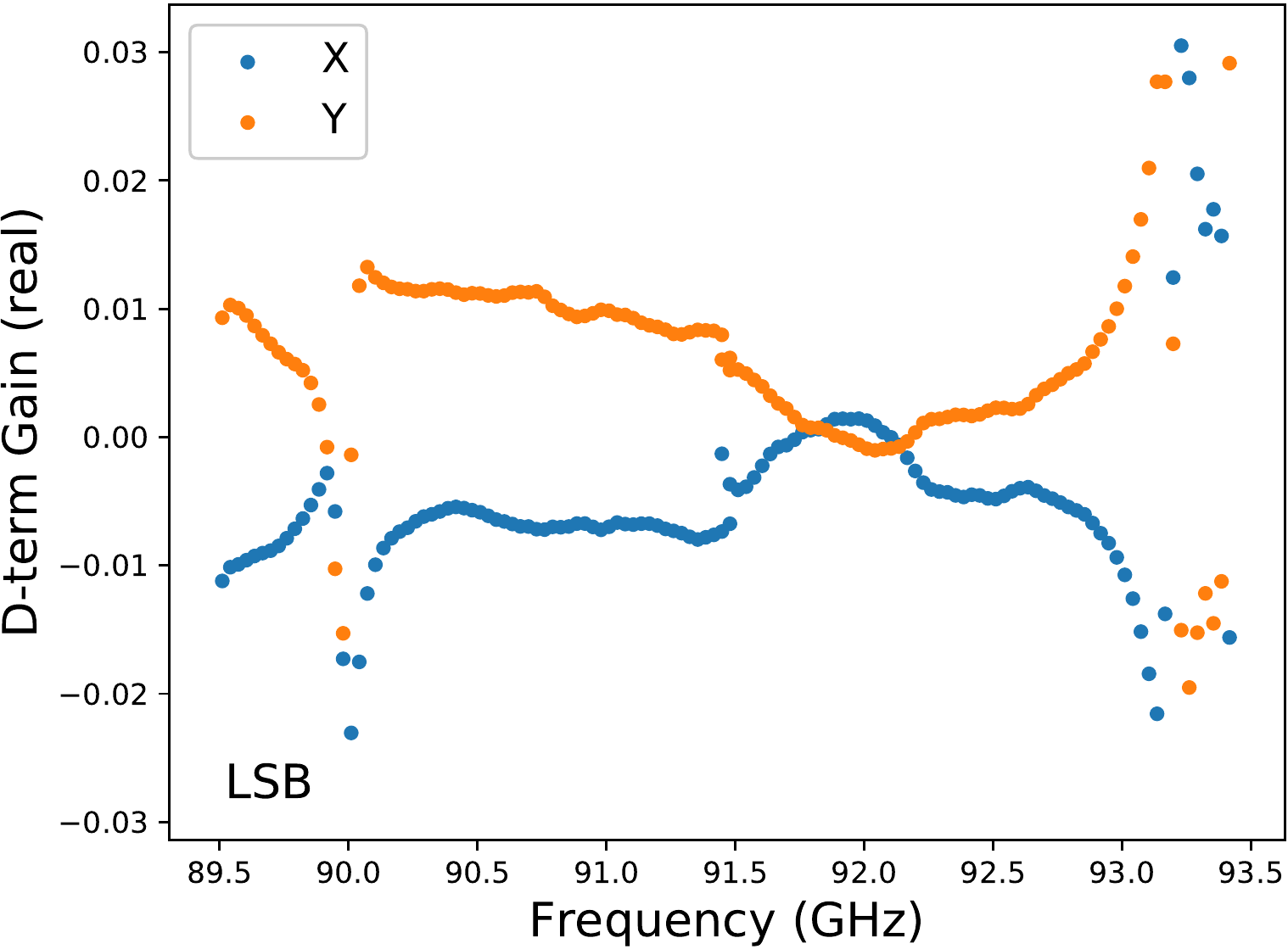}  &
   \includegraphics[width=0.31\textwidth]{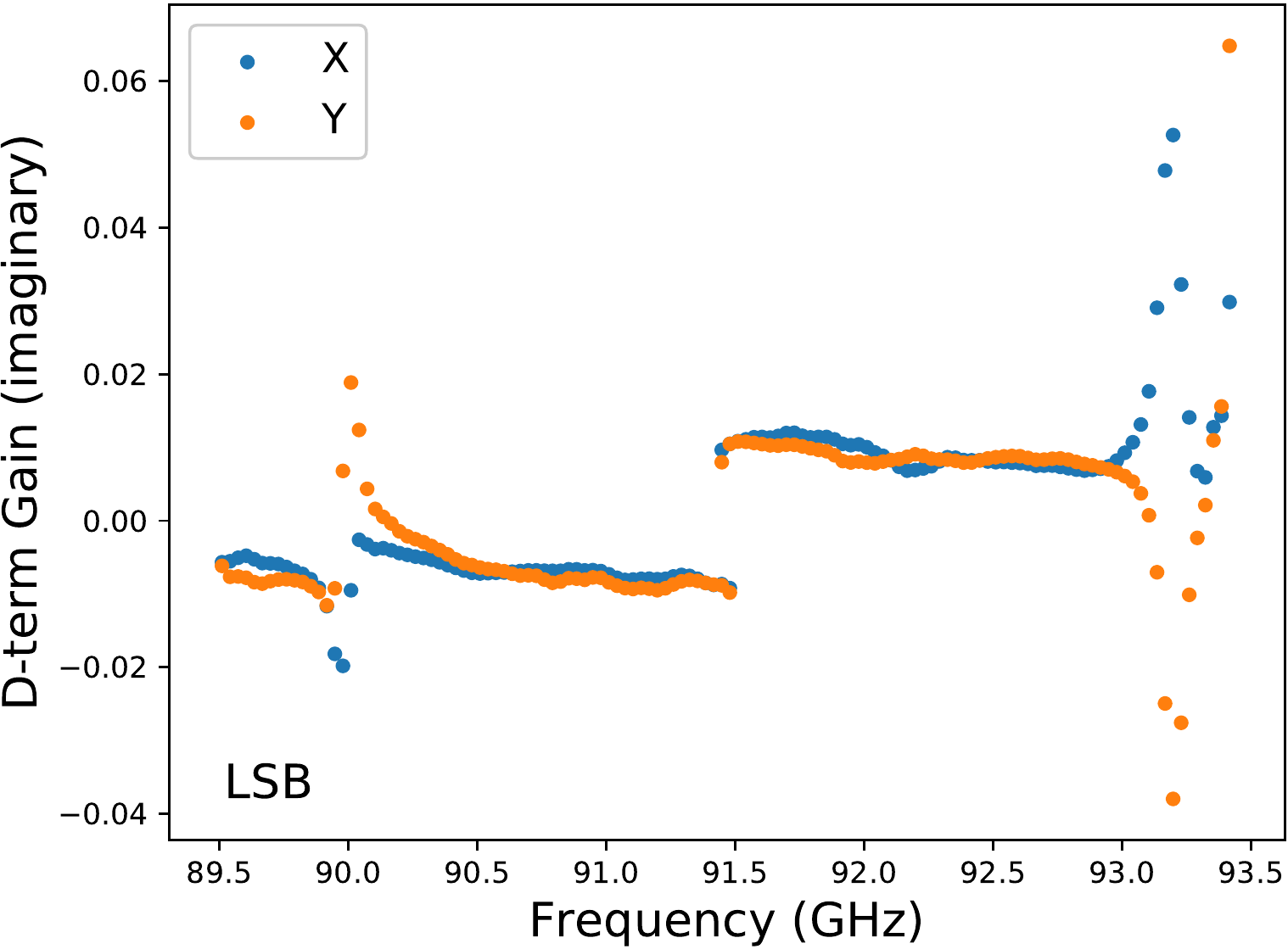}  &
   \includegraphics[width=0.31\textwidth]{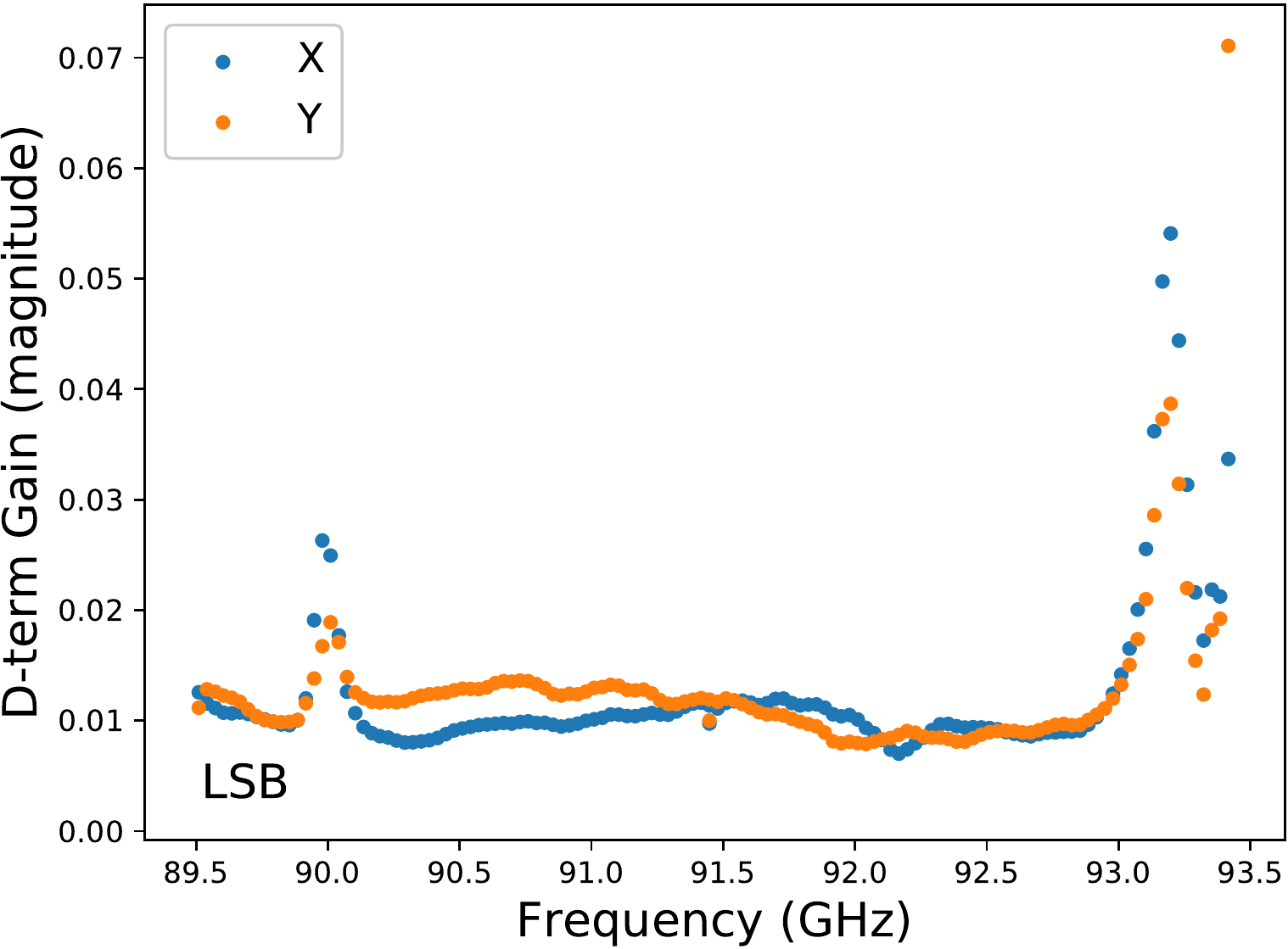}
 \end{tabular}  
  \caption{Real (left column), imaginary (center column), and magnitude (right column) of the 
derived complex polarization leakage (``$D$-terms'') for antenna DA50 in the upper sideband (USB; 
upper row) and lower sideband (LSB; lower row) for both $X$ (blue) and $Y$ (orange) polarizations
for a reduction using DA64 as reference antenna. 
The LSB leakage exhibits a peak at $\approx90$~GHz and a large spike above $\approx93$~GHz, while 
the USB leakage has a quasi-periodic structure. These structures are robust to choice of reference 
antenna used in the polarization calibration (Fig.~\ref{fig:refantcomp_dterms}). $D$-term solutions 
for the other antennas exhibit similar trends. }
\label{fig:dterms}
\end{figure*}

\subsection{Detailed data analysis}
\label{text:reduction}
Given the apparent instability of polarization properties of the target and phase calibrator with 
both time and frequency in the QA2 results, we perform a full independent reduction of the data. 
We import the raw ASDM datasets into CASA, followed by flagging of non-interferometric (e.g. 
pointing, atmospheric calibration, and sideband ratio) data. We apply the system temperature 
(Tsys) and water vapor radiometer (wvr) calibrations to the data, and concatenate the three 
executions of the scheduling block (SB) into a single CASA measurement set. 

We perform interferometric and polarization calibration using standard techniques, beginning with 
deriving the bandpass phase and amplitude calibration, in that order. We use DV06 as reference 
antenna, and validate our calibration by repeating the entire analysis separately using a nearby 
antenna with a different architecture, DA64. For the polarization calibration, we first derive the 
complex gain solutions on the polarization calibrator, and then derive an \textit{a priori} 
estimate 
of
its Stokes $Q$ and $U$ from the ratio of complex gains 
using the python utility \texttt{qufromgain} from the ALMA polarization 
helpers module (\texttt{almapolhelpers.py}; see CASA documentation for details). The parallactic 
angle of the polarization calibrator decreases from $\approx230^\circ$ to $\approx130^\circ$ over 
the course of the observations, providing adequate coverage for disentangling the source and 
instrumental polarization. 
The derived fractional $Q$ and $U$ values for the polarization calibrator are consistent across all 
four spectral windows and across the use of the two different reference antennas, although we note 
that using DV06 yields a lower estimated uncertainty on $Q$ and $U$ (Table~\ref{tab:polcal}). We 
note that these are fractional polarization values, since they were derived assuming unity Stokes 
$I$.

To derive the cross-hand delays, we use scan 61 on the polarization calibrator as the scan 
with the strongest polarization signal, selected based on a plot of the complex polarization ratio 
for this calibrator as a function of time\footnote{See 
\url{https://casaguides.nrao.edu/index.php/3C286_Polarization} for a description of this process.}. 
We next solve for the $XY$ phase of the reference antenna, the channel-averaged polarization of 
the polarization calibrator, and the instrumental polarization using the XYf+QU mode in CASA's 
\texttt{gaincal} task. The net instrumental polarization averaged across all baselines 
(as reported by \texttt{gaincal})
varies from $\approx0.06\%$ to $\approx0.2\%$ over the four spectral windows. We resolve the $QU$ 
phase ambiguity with the python utility \texttt{xyamb} using the fractional $Q$ and $U$ derived 
earlier. We list the final derived values for the fractional polarization of the polarization 
calibrator in Table~\ref{tab:polcal}. 

\begin{figure}
  \centering
 \begin{tabular}{cc}
   \includegraphics[width=0.22\textwidth]{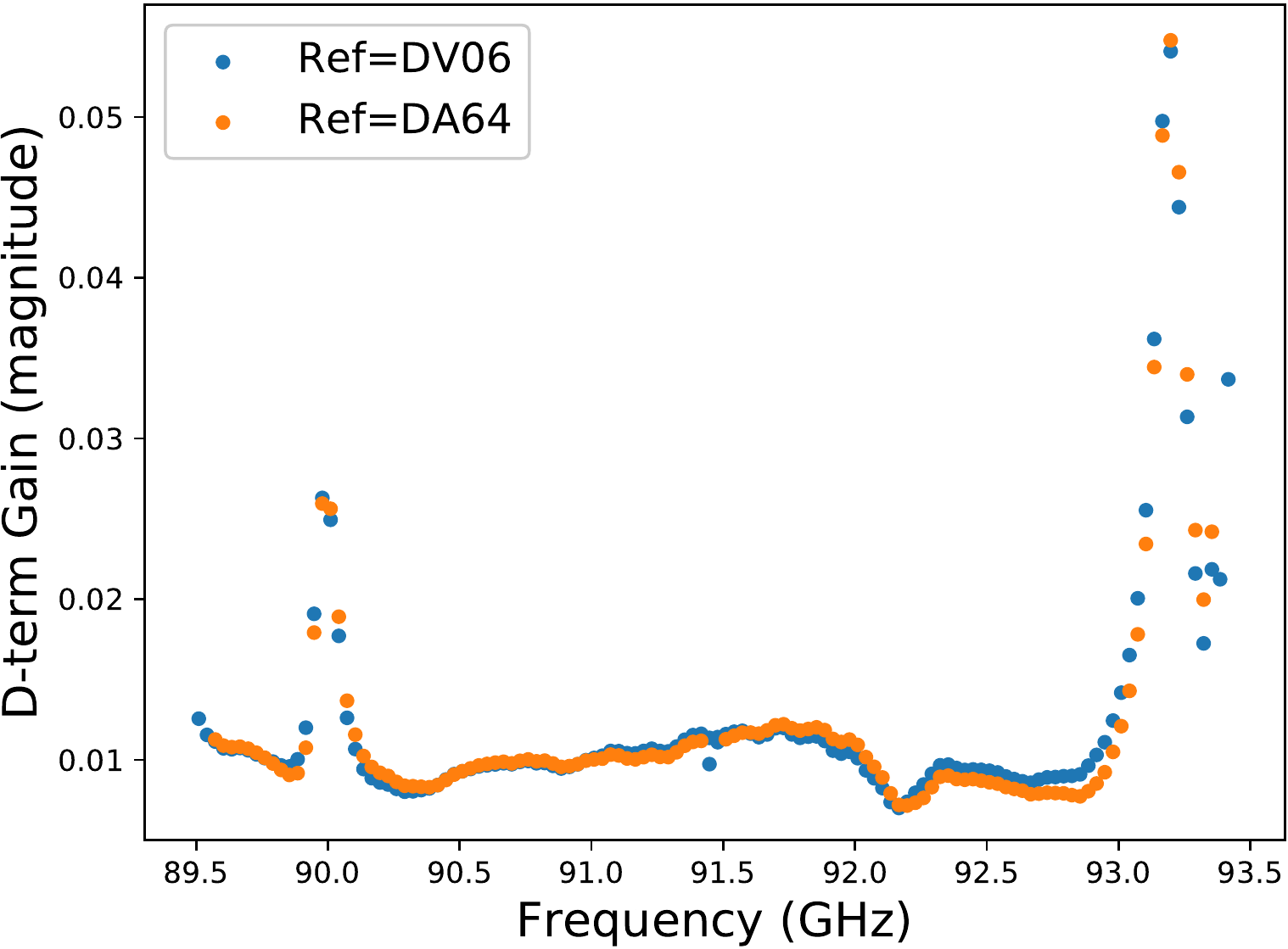}  &
   \includegraphics[width=0.22\textwidth]{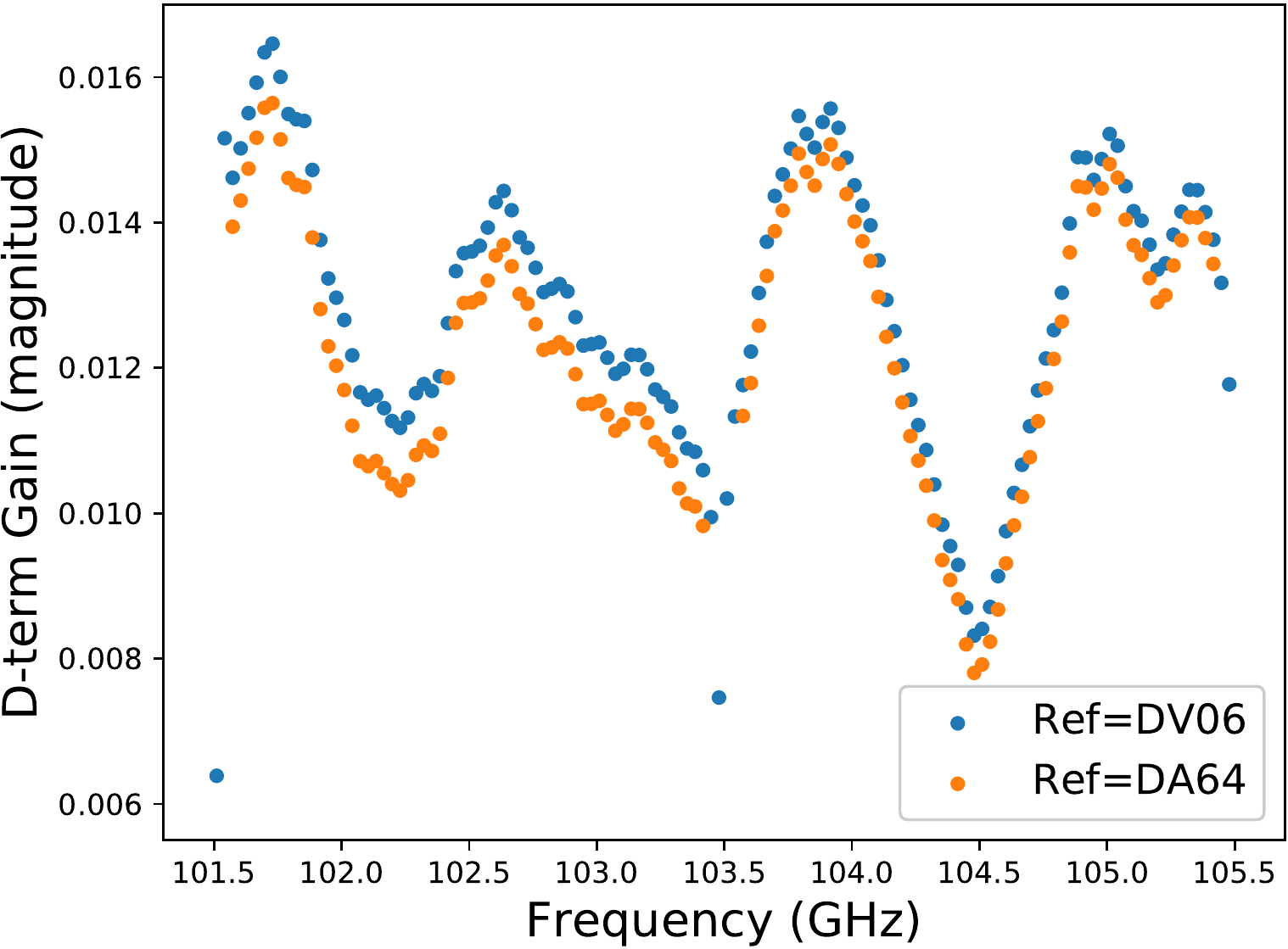} 
 \end{tabular}  
  \caption{Magnitude of the polarization leakage for antenna DA50 in the lower sideband (left) and 
upper sideband (right) derived using independent reductions with two different reference antennas, 
DV06 (blue) and DA64 (orange). The derived $D$-terms are robust to choice of reference antenna 
used. $D$-term solutions for the other antennas exhibit similar trends. }
\label{fig:refantcomp_dterms}
\end{figure}

We use these derived polarization properties to refine the complex gain solution on the 
polarization calibrator. We run \texttt{qufromgain} again on the resulting calibration table, which 
yields a residual polarization statistically indistinguishable from zero, and demonstrates that 
the source polarization has been successfully removed from the gain solutions. However, we 
note that the final residual polarization, determined by running \texttt{qufromgain} on the 
calibrated calibrator data, is $\approx0.1\%$ (Table~\ref{tab:polcal}), suggesting that the minimum 
systematic 
uncertainty in polarization measurements from this dataset is at least of this order. Antenna DV22 
exhibits large ($\approx10\%$) residual cross-hand polarization amplitude gain ratios in the 
91.5--93.5~GHz spectral window, and we flag that antenna in that spectral window before proceeding.

Finally, we solve for the polarization leakage (antenna ``$D$ terms'') using \texttt{polcal}. The 
derived leakage terms exhibit a strong increase up to $\approx7\%$ for several antennas at the 
upper edge ($\gtrsim93$~GHz) of the LSB, in addition to a weaker peak at $\approx 90$~GHz 
(Fig.~\ref{fig:dterms}). This behavior is seen in both reductions\footnote{We also tested our 
analysis by flagging these channels, but this did not significantly change the results of the 
subsequent imaging.}, i.e., independent of the reference antenna used for calibration 
(Fig.~\ref{fig:refantcomp_dterms}). The leakage appears lower and more consistent across channels 
in the USB, but does exhibit a quasi-periodic structure, as previously also noted in the 3C286 
Science Verification data of ALMA Band 6 polarization observations 
\citep{nnp+16}.

We set the flux density of J1127-1857 using measurements near the time of the GRB 
observations listed the ALMA calibrator catalog, from which we derive a spectral index of 
$\beta=-0.51\pm0.01$ and a flux density of $\approx1.05$~Jy at a reference frequency of 91.5~GHz. 
The derived flux density of the polarization calibrator (J1256-0547) is $12.4871\pm0.0009$~Jy 
at the band center reference 
frequency of 97.287~GHz with a spectral index of $\beta=-0.528\pm0.001$, and that of the gain
calibrator (J1130-1449) is 
$0.7968\pm0.0004$~Jy with a spectral index of $\beta=-0.987\pm0.007$. We complete the calibration 
by deriving and applying standard interferometric complex antenna gain solutions using the 
interleaved observations of J1130-1449. 

\begin{deluxetable*}{cllrrrrrrr}
 \tabletypesize{\footnotesize}
 \tablecolumns{11}
  \tablecaption{Impact of self-calibration on ALMA Band 3 (97.5 GHz) Polarization Observations of 
GRB~171205A}
  \tablehead{
   \colhead{Sideband} &
   \colhead{Selfcal} &
   \colhead{Selfcal} &   
   \colhead{$I$} &
   \colhead{$I_{\rm rms}$} &
   \colhead{$Q$} &
   \colhead{$U$} &   
   \colhead{$V$} &   
   \colhead{$P$} \\
   &
   \colhead{Type} &
   \colhead{Interval\tablenotemark{a}} &
   \colhead{(mJy)} &
   \colhead{($\mu$Jy)} &
   \colhead{($\mu$Jy)} &
   \colhead{($\mu$Jy)} &
   \colhead{($\mu$Jy)} &   
   \colhead{($\mu$Jy)} 
    }
 \startdata 

LSB & None & \ldots & $31.68\pm0.11$ & 38.0 & $-3.2\pm10.6$ & $50.8\pm11.2$ & $-65.2\pm10.4$ & 
$69.6\pm15.4$ 
\\
LSB & phase only & 10~min & $32.68\pm0.03$ & 15.8 & $-4.2\pm10.7$ & $53.9\pm11.2$ & $-67.2\pm10.3$ 
& $72.9\pm15.4$ \\
LSB & phase only & 2~min & $33.13\pm0.02$ & 11.4 &  $-4.3\pm10.8$ & $55.5\pm11.4$ & $-68.4\pm10.4$ 
& $73.9\pm15.5$ \\ 
LSB & amp \& phase & 20~min & $33.02\pm0.03$ & 11.2 & $14.4\pm10.6$ & $72.8\pm11.4$ & 
$-73.1\pm10.3$ & $86.7\pm15.6$ \\
LSB & amp \& phase & 30~s\tablenotemark{b} & $33.11\pm0.03$ & 10.5 &  $4.6\pm10.1$ & $72.4\pm11.6$ 
& $-70.4\pm10.4$ 
& $87.1\pm15.2$ \\
\hline
USB & None & \ldots & $29.96\pm0.12$ & 35.3 &  $-54.0\pm10.8$ & $-46.3\pm10.3$ & $-69.8\pm10.4$ & 
$77.5\pm14.9$ \\
USB & phase only & 10~min & $31.18\pm0.04$ & 14.7 &  $-55.5\pm10.8$ & $-44.4\pm10.5$ & 
$-75.0\pm10.5$ & 
$75.9\pm15.0$ \\
USB & phase only & 2~min & $31.79\pm0.03$ & 12.2 &  $-57.6\pm10.8$ & $-45.8\pm10.6$ & 
$-76.9\pm10.7$ & 
$78.6\pm15.1$ \\
USB & amp \& phase & 20~min & $32.00\pm0.03$ & 11.3 &$-57.2\pm10.6$ &$-44.9\pm10.4$ & 
$-77.5\pm10.5$ & $77.4\pm14.9$ 
\\
USB & amp \& phase & 30~s\tablenotemark{b} & $32.08\pm0.03$ & 10.6 & $-7.1\pm10.0$ &$-46.1\pm10.5$ 
& 
$-77.8\pm10.5$ 
& $51.5\pm14.5$\\
\hline
All &None & \ldots  & $30.77\pm0.09$ & 29.3 & $-20.0\pm7.8$ &$13.6\pm8.1$ & $-70.0\pm7.6$ & 
$37.8\pm11.2$\\
All &amp \& phase & 20~min & $32.44\pm0.03$ & 7.7 & $-31.5\pm7.3$ &$9.6\pm7.7$ & $-73.4\pm7.3$ & 
$41.7\pm10.6$
\enddata
 \label{tab:scalfluxes}
 \tablenotetext{a}{Cross-hand phase fixed for 20-minute solutions, and left free for 30-second 
solutions.
{\tablenotemark{b}For comparison with the analysis of \cite{uth+19}.} }
\end{deluxetable*}

\subsection{Imaging}
We combine and image the calibrated measurement set using \texttt{tclean} in CASA with a robust 
parameter of 0.0 and one Taylor term (i.e.~nterms=1).
The clean beam is $0\farcs28$\,$\times$\,$0\farcs20$ at a position angle of 
$85^\circ$ 
The afterglow is well-detected with 
a flux density of $30.8\pm0.1$~mJy, measured with 
a point source model using \texttt{imfit} in CASA. No significant polarization signal is detected 
at the position of the afterglow in Stokes $Q$, $U$, or in the $P$ image. Our initial estimate for 
the point source flux density is $\approx4\%$ lower than the self-calibrated and sideband-combined 
flux density reported by \cite{uth+19}. However, we caution against a direct flux comparison, since 
\cite{uth+19} do not report the flux density or spectral properties of the flux calibrator that 
they assumed for the analysis. 

We note the presence of significant ($\approx3\%$) cleaning residuals in the Stokes $I$ image, both 
for the GRB and the phase calibrator, indicating residual calibration errors, potentially due to 
atmospheric phase decoherence\footnote{For reference, the phase calibrator J1130-1449 is $5{\fdg}4$
from the GRB position.}. 
We correct for these by performing two rounds of phase-only 
self-calibration with solution intervals of 10~min and 2~min on both the GRB afterglow and phase 
calibrator data. We split\footnote{We also average the data to a 6s integration time and decimate 
by 
2 channels in order to reduce the data volume. The resulting total beam smearing across the 
$25\arcsec\times25\arcsec$ image is $\approx0.02\arcsec$, which is a fraction of the $0.05\arcsec$ 
cell size, much smaller than the synthesized beam, and negligible for a point source near the field 
center.} the data into upper and lower sidebands for this step in order to reduce the fractional 
bandwidth from $\approx16\%$ for the full dataset to $\approx4\%$ per sideband, and thus minimize 
the effect of the frequency structure of the source on the calibration solutions. This is 
especially 
important for the calibrator, which exhibits a fitted spectral index (from the gain solutions) of 
$\approx-0.5$, and thus a potential variation in Stokes $I$ intensity of $\approx8\%$ across the 
ALMA band. We solve for a single gain solution for both polarizations (gain mode `T') using 
\texttt{gaincal} in CASA, in order to avoid introducing a phase offset between the $X$ and $Y$  
polarizations. Additionally, we set the reference antenna mode to \texttt{strict} to enforce the 
use of a single reference antenna during the self-calibration. We continue the use of the same 
reference antenna for self-calibration as that employed during the earlier calibration steps. 

We fit the Stokes $I$ image with a point source model using CASA \texttt{imfit}, followed by fits 
to the $QUVP$ images with the position and beam parameters fixed to that derived from the Stokes 
$I$ image. We perform point source fits at each step during the phase-only self-calibration, and 
present these, {together with the Stokes $I$ map rms,} in Table~\ref{tab:scalfluxes} for 
reference. 
The phase self-calibration reveals low-level ($\approx2\%$) symmetric residuals indicative of 
amplitude-based errors. We, therefore, perform one round of amplitude and phase self calibration, 
applying the pre-derived phase solutions on the fly. Since amplitude self-calibration is a less 
well constrained problem, we solve for one solution per 20~min, for a total of 14 solutions per 
polarization per antenna. The derived amplitude solutions exhibit moderate ($\approx10\%$) 
variability with time, but the flux density scale remains stable under amplitude self calibration
(Table~\ref{tab:scalfluxes}). 

We find marginally decreasing residuals with shorter solution 
intervals; however, amplitude self-calibration at intervals shorter than 20~min do not improve the 
signal-to-noise further. In particular, a 30~s amplitude and phase self-calibration with gains 
for both polarizations solved independently as performed in \cite{uth+19} does not yield a 
measureable improvement in signal-to-noise in Stokes $I$ (Table~\ref{tab:scalfluxes}). Furthermore, 
these symmetric residuals are not completely removable even with 30~s amplitude and phase 
self-calibration, suggesting that the errors may be baseline-based, rather than antenna-based. 
Finally, we note that the minimum theoretical solution interval for self-calibration ($t_{\rm 
solint}$), 
which is given by 
\begin{equation}
\frac{I_{\rm peak}}{\sigma_I} > 3\sqrt{N-3}\sqrt{\frac{t_{\rm int}}{t_{\rm solint}}},
\end{equation}
where is $t_{\rm int}\approx9.4\times10^3$~s is the total integration time on source,
$N=43$ is the number of antennas in the array, $I_{\rm peak}\approx30$~mJy is the peak intensity
of the source used for self-calibration, 
and $\sigma_{\rm I}\approx0.1~$mJy is the off-source image rms prior to self-calibration,
yields $t_{\rm solint}\gtrsim35$~s. Thus, the 30~s solution interval used by \cite{uth+19} is 
shorter than the minimum possible $t_{\rm solint}$ where stable solutions may be expected. 

We perform point source fits on our final images (amplitude and phase self-calibrated to 20 
min, with the cross-hand phase fixed), as well as on images made using 30 s amplitude and phase 
self calibration, where the $X$ and $Y$ gains were allowed to vary independently. We find that
reducing the solution interval and fitting the cross-hand phase yields only a marginal increase in 
Stokes $I$ flux density, from $33.02\pm0.03$~mJy to $33.11\pm0.03$~mJy in the lower sideband, and 
from $32.00\pm0.03$ to $32.08\pm0.03$ in the upper sideband.
For comparison, we also combine the self-calibrated sideband-separated $uv$-data into a single
measurement set, and image the entire 4~GHz dataset simultaneously.
Except for Stokes $U$ in the LSB, no significant ($\gtrsim5\sigma$) emission is visible in the 
Stokes $QUP$ images. On the other hand, significant ($\approx10\sigma$) 
circular polarization again appears at the $\approx0.23\%$ level. 

\begin{figure*}
 \centering
 \begin{tabular}{ccc}
  \includegraphics[width=0.31\textwidth]{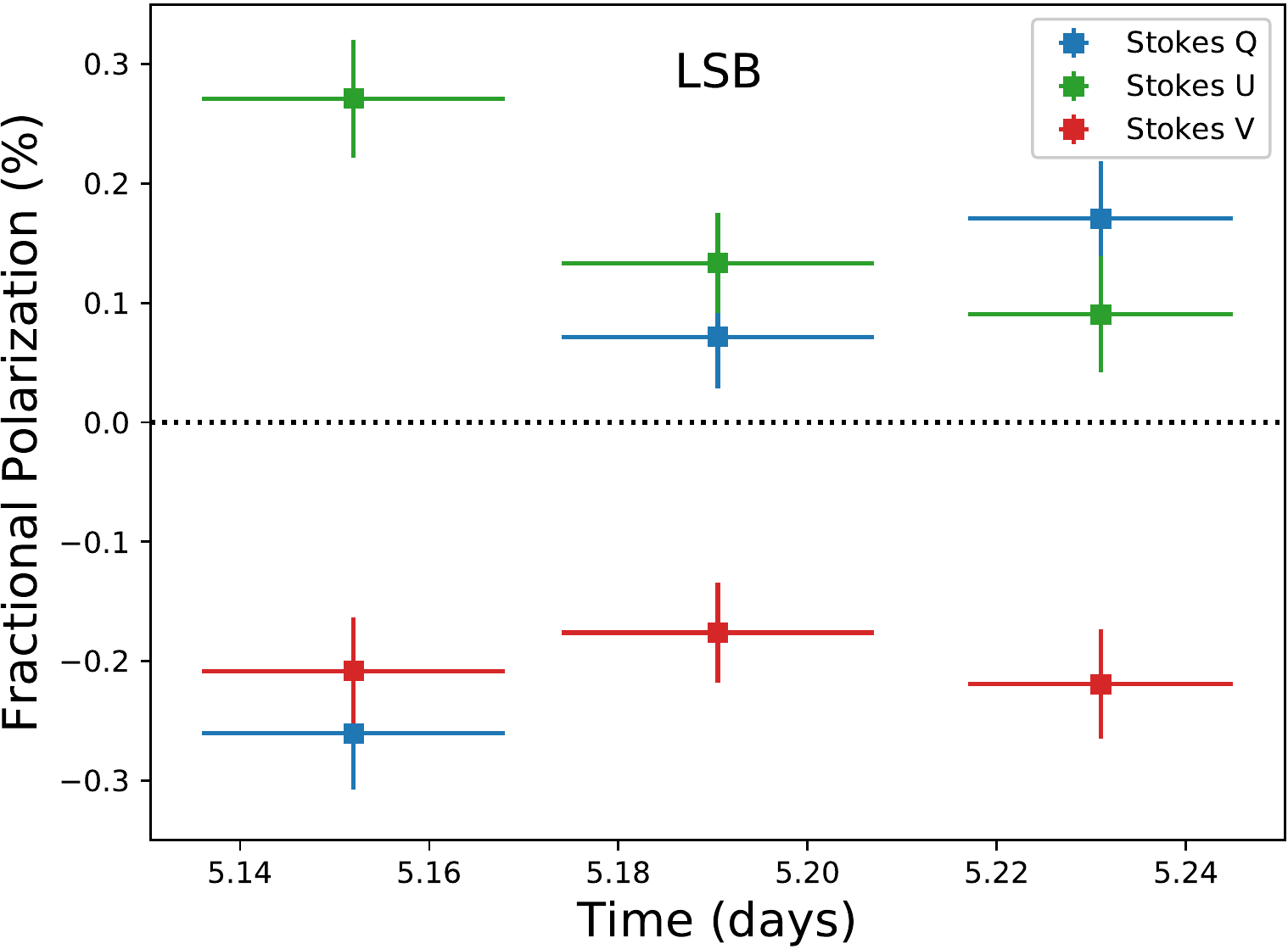} &
  \includegraphics[width=0.31\textwidth]{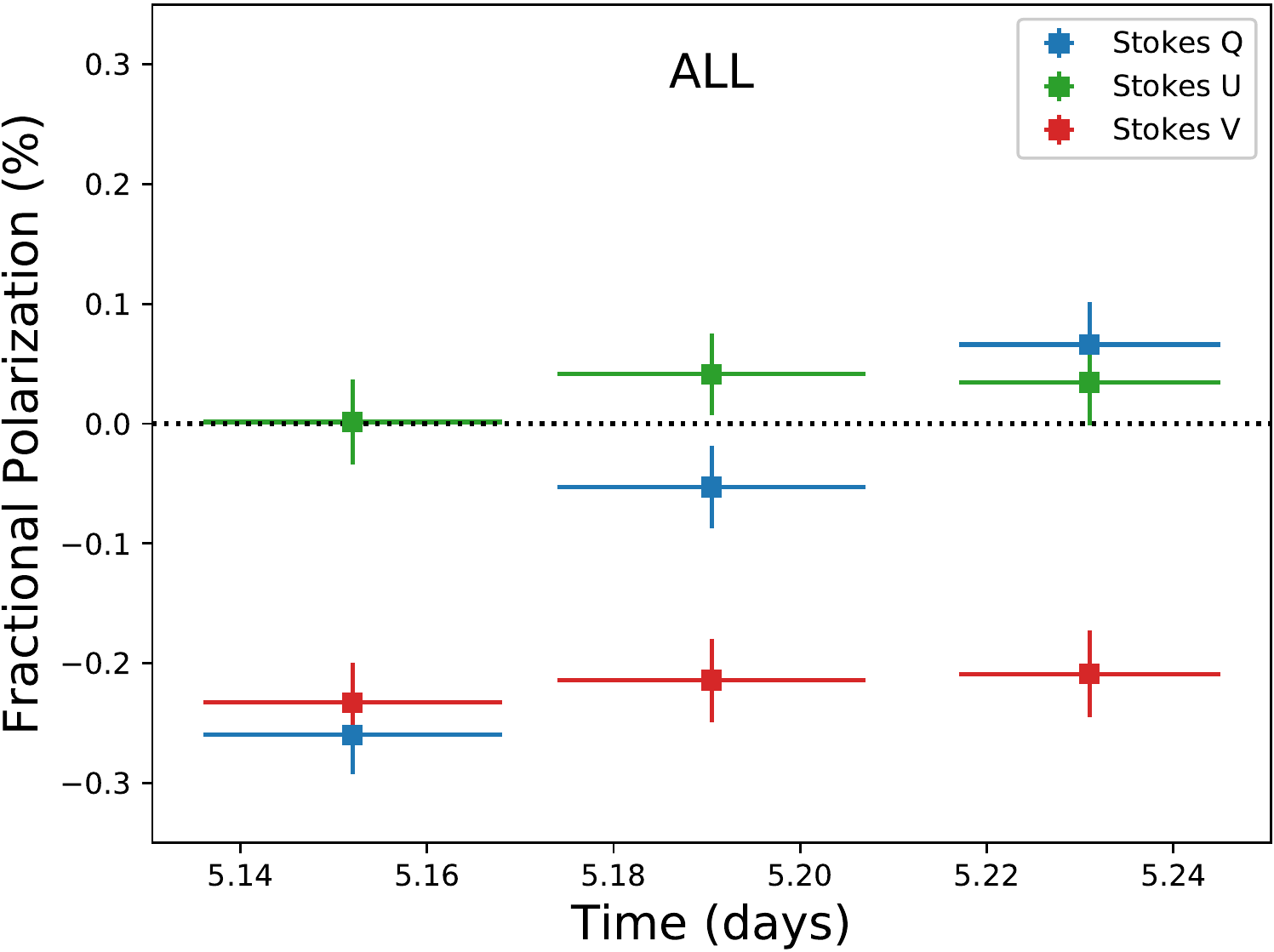} &
  \includegraphics[width=0.31\textwidth]{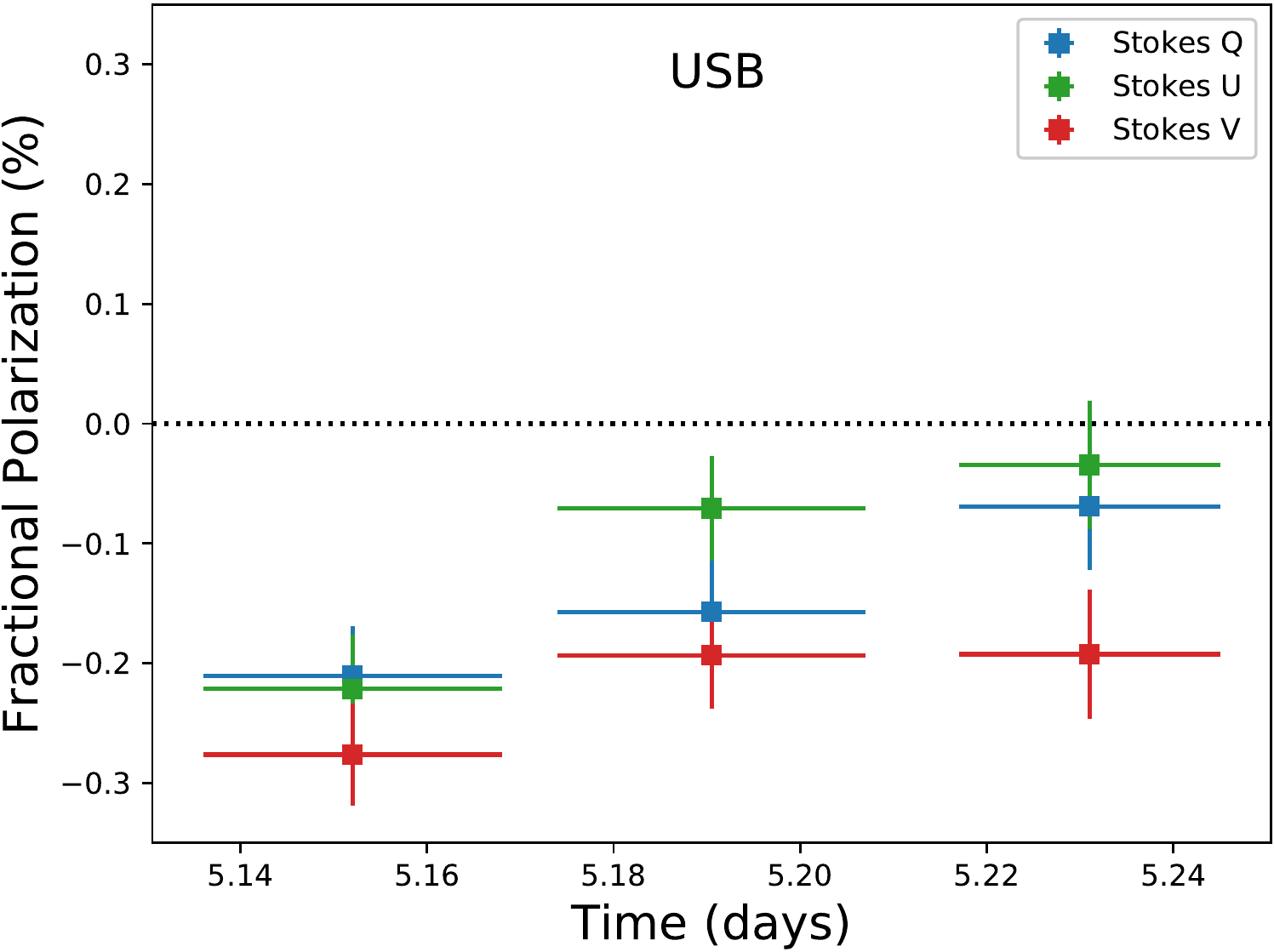} 
 \end{tabular}
 \caption{Fractional Stokes $QUV$ measurements for GRB~171205A (using self-calibrated data) as a 
function of time relative to the \textit{Swift} trigger time (center), divided into LSB (left) 
and USB (right). Each time bin corresponds to one execution of the scheduling block. The fractional 
uncertainty from the Stokes $I$ measurement is two orders of magnitude smaller than that from the 
$QUV$ measurements, and is thus ignored. Error bars in the time direction correspond to the span 
of the data imaged.}
\label{fig:varQUV_GRB}
\end{figure*}

\begin{figure*}
 \centering
 \begin{tabular}{ccc}
  \includegraphics[width=0.31\textwidth]{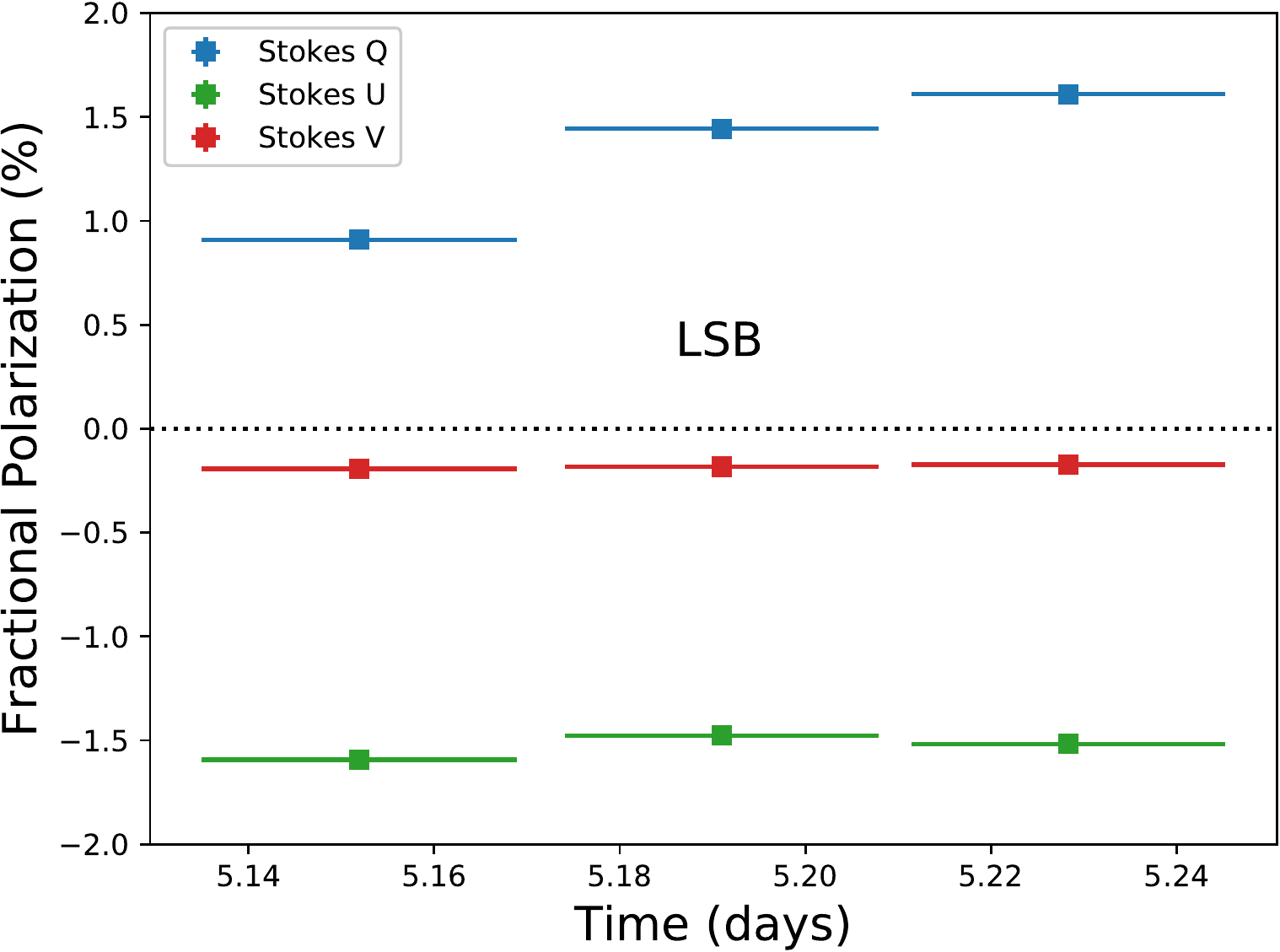} &
  \includegraphics[width=0.31\textwidth]{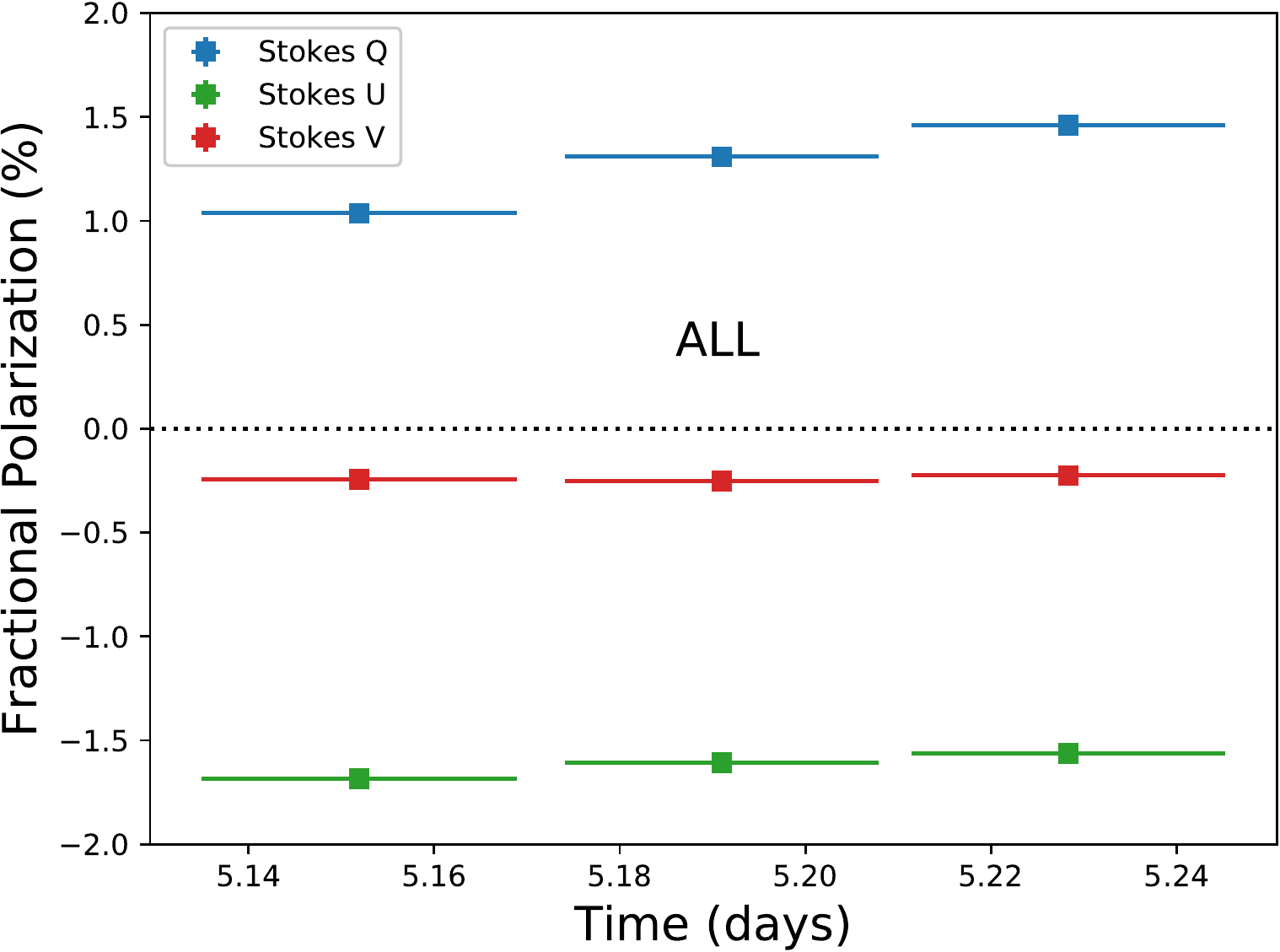} &
  \includegraphics[width=0.31\textwidth]{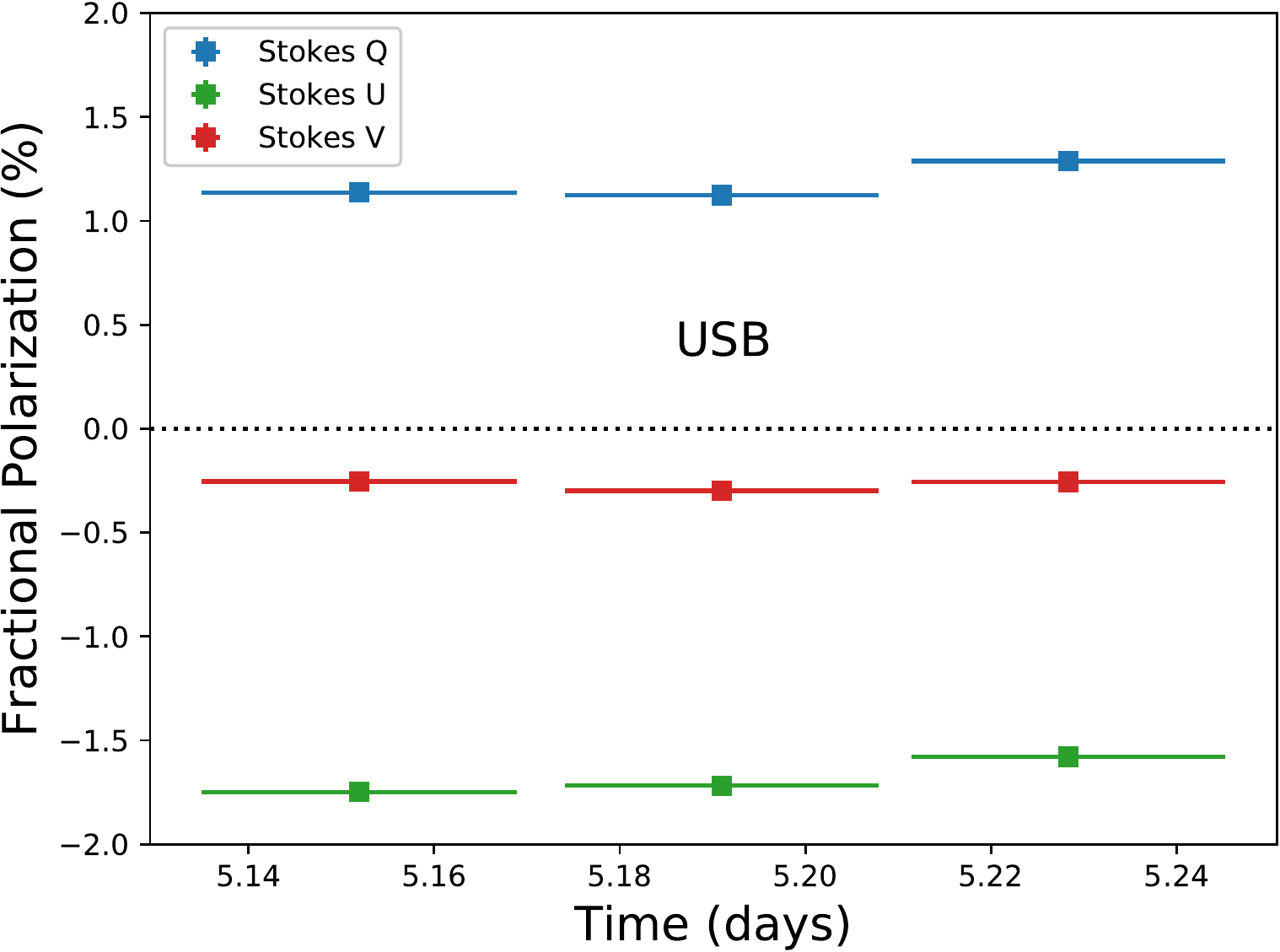} \\
  \includegraphics[width=0.31\textwidth]{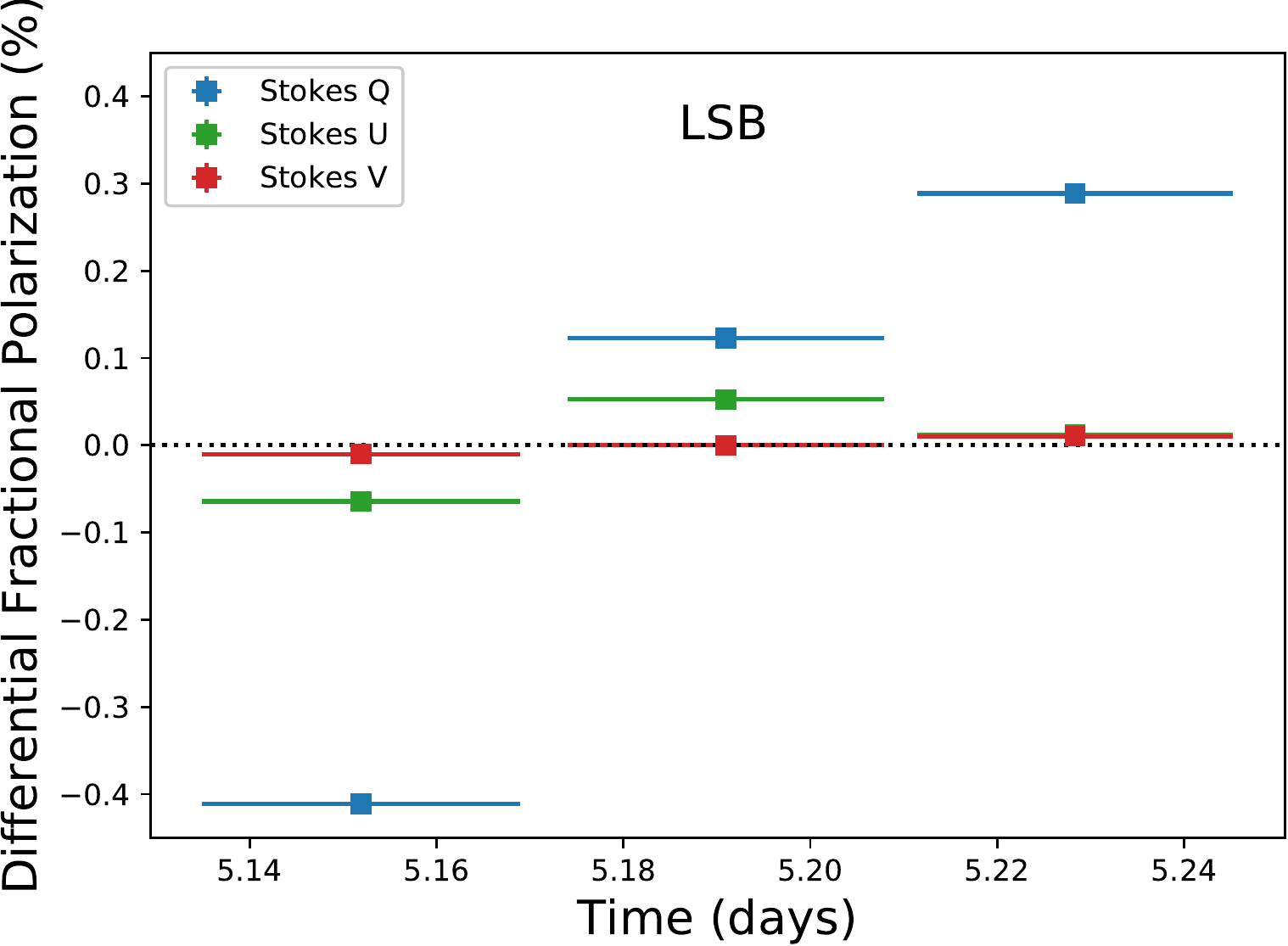} &
  \includegraphics[width=0.31\textwidth]{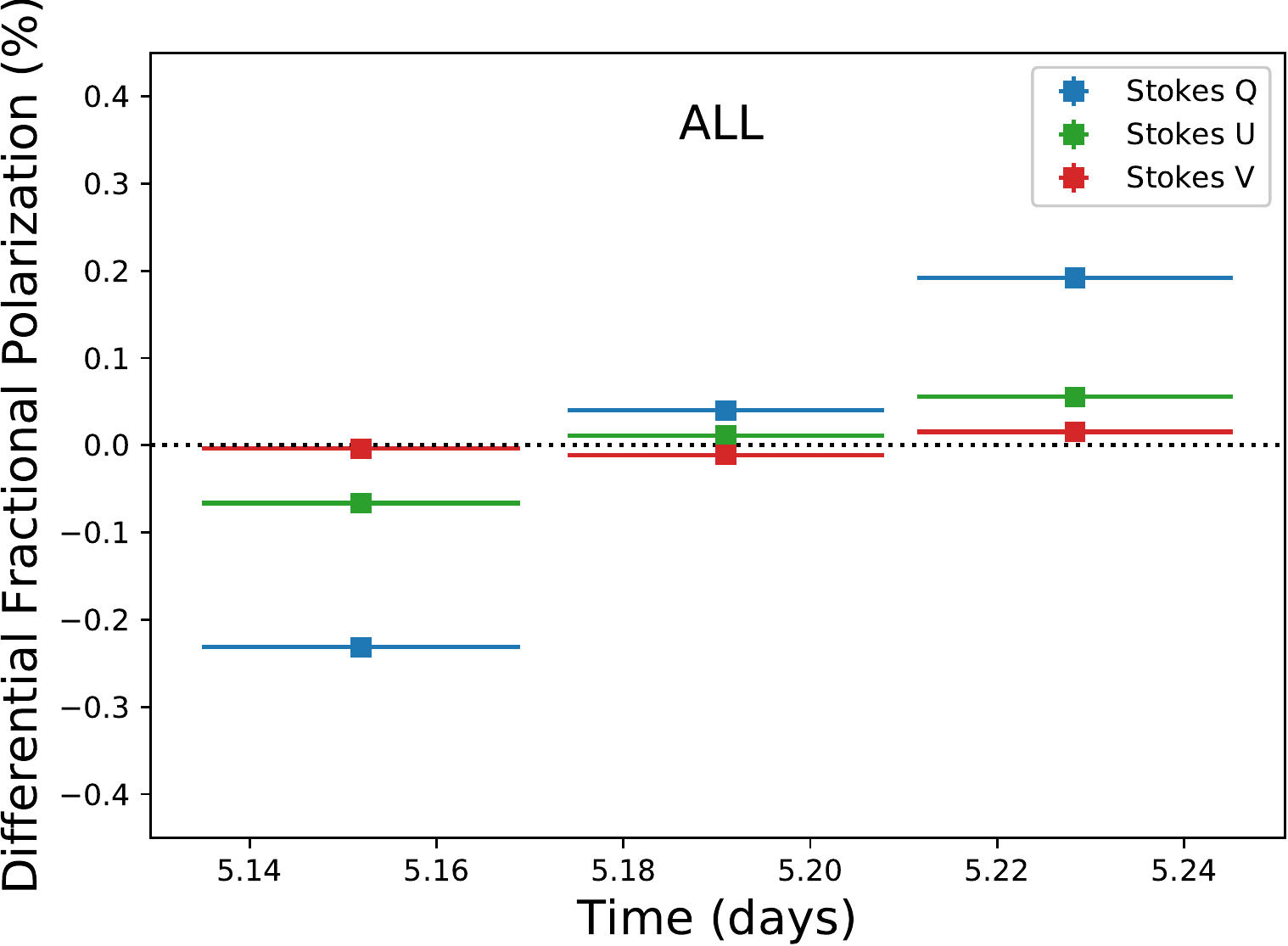} &
  \includegraphics[width=0.31\textwidth]{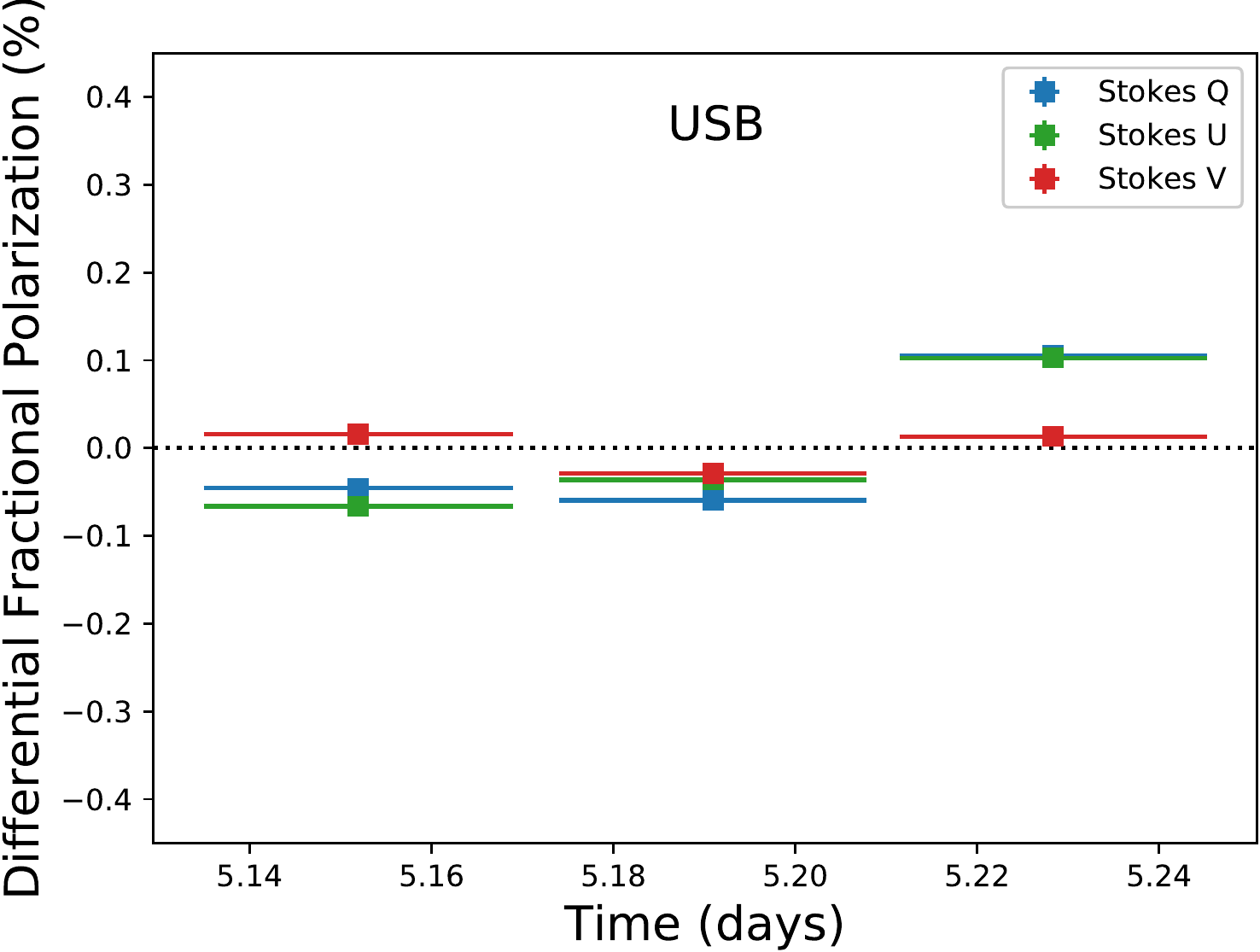} 
 \end{tabular}
 \caption{Upper panels: same as Fig.~\ref{fig:varQUV_GRB}, but for the gain calibrator J1130-1449. 
The (statistical) uncertainties on each point are typically smaller than the thickness of the line 
used to plot the horizontal error bar. We find a non-zero Stokes $V$, as well as significant 
evolution of Stokes $QUV$ with time. The latter effect is clearer with the respective mean values 
of 
$QUV$ removed (lower panels; mean subtracted independently for each subplot and polarization). All 
plots in the same row are on the same scale.}
\label{fig:varQUV_GCal}
\end{figure*}

\begin{deluxetable*}{ccrcrrrrrrr}
 \tabletypesize{\footnotesize}
 \tablecolumns{11}
  \tablecaption{Stability of ALMA Band 3 Polarization Observations to Time and Frequency Slicing}
  \tablehead{
   \colhead{Target} &   
   \colhead{Sideband} &   
   \colhead{Frequency} &   
   \colhead{SB\tablenotemark{a}} &
   \colhead{$I$} &
   \colhead{$I_{\rm rms}$} &
   \colhead{$Q$} &
   \colhead{$U$} &   
   \colhead{$V$} &   
   \colhead{$P$} \\
   &  &   
   \colhead{(GHz)} &
   \colhead{Execution} &
   \colhead{(mJy)} &
   \colhead{($\mu$Jy)} &
   \colhead{($\mu$Jy)} &
   \colhead{($\mu$Jy)} &
   \colhead{($\mu$Jy)} &   
   \colhead{($\mu$Jy)} 
    }
 \startdata 
  GRB~171205A & LSB &  91.463 & 1 & $33.01\pm0.03$ & 16.6 & $-86.0\pm15.6$ & $89.5\pm16.3$ & 
$-68.7\pm14.8$ 
& $140.5\pm22.4$\\
  GRB~171205A & LSB &  91.463 & 2 & $33.04\pm0.04$ & 16.1 & $23.7\pm14.3$ & $44.0\pm13.8$ & 
$-58.2\pm13.8$ 
& $67.0\pm19.8$\\
  GRB~171205A & LSB &  91.463 & 3 & $33.07\pm0.04$ & 17.4 & $56.4\pm15.8$ & $29.9\pm16.1$ & 
$-72.5\pm15.2$ 
& $81.1\pm22.5$\\
  GRB~171205A & USB & 103.495 & 1 & $32.00\pm0.03$ & 17.4 & $-67.4\pm13.4$ & $-70.9\pm14.4$ & 
$-88.4\pm13.5$ 
& $105.3\pm19.6$ \\
  GRB~171205A & USB & 103.495 & 2 & $32.01\pm0.04$ & 16.9 & $-50.3\pm13.6$ & $-22.6\pm14.0$ & 
$-61.9\pm14.2$ 
& $61.2\pm19.6$\\
  GRB~171205A & USB & 103.495 & 3 & $32.08\pm0.04$ & 18.3 & $-22.1\pm17.0$ & $-11.0\pm17.1$ & 
$-61.7\pm17.3$ 
& $50.4\pm24.1$ \\
  GRB~171205A & All &  97.496 & 1 & $32.46\pm0.03$ & 12.5 & $-84.3\pm10.7$ & $0.05\pm11.6$ & 
$-75.6\pm10.8$ 
& $91.3\pm15.7$ \\
  GRB~171205A & All &  97.496 & 2 & $32.52\pm0.03$ & 12.2 & $-17.2\pm11.3$ & $13.5\pm11.1$ & 
$-69.7\pm11.3$ 
& $34.5\pm15.8$ \\
  GRB~171205A & All &  97.496 & 3 & $32.54\pm0.03$ & 13.8 &  $21.5\pm11.6$ & $11.2\pm11.7$ & 
$-68.0\pm11.9$ 
& $44.1\pm16.4$ \vspace{0.05in}\\
\hline\rule{0pt}{2.6ex}
       &     &         &   &       &  & $Q$ (mJy) & $U$ (mJy) & $V$ (mJy) & $P$ (mJy) \\
  Gain Calibrator & LSB &  91.463 & 1 & $847.29\pm0.08$ & 69.8 & $7.71\pm0.07$ & $-13.50\pm0.09$ & 
$-1.63\pm0.05$ & $15.5\pm0.1$ \\
  Gain Calibrator & LSB &  91.463 & 2 & $847.61\pm0.08$ & 69.4 & $12.20\pm0.10$ & $-12.52\pm0.09$ & 
$-1.55\pm0.05$ & $17.5\pm0.1$ \\
  Gain Calibrator & LSB &  91.463 & 3 & $847.83\pm0.08$ & 68.0 & $13.60\pm0.10$ & $-12.86\pm0.07$ & 
$-1.46\pm0.05$ & $18.8\pm0.1$ \\
  Gain Calibrator & USB & 103.495 & 1 & $749.19\pm0.07$ & 61.3 & $8.52\pm0.06$ & $-13.10\pm0.08$ & 
$-1.90\pm0.05$ & $15.6\pm0.1$ \\
  Gain Calibrator & USB & 103.495 & 2 & $749.17\pm0.06$ & 66.2 & $8.42\pm0.09$ & $-12.87\pm0.08$ & 
$-2.23\pm0.05$ & $15.4\pm0.1$ \\
  Gain Calibrator & USB & 103.495 & 3 & $749.32\pm0.06$ & 62.5 & $9.70\pm0.10$ & $-11.8\pm0.06$ & 
$-1.92\pm0.06$ & $15.3\pm0.1$ \\
  Gain Calibrator & All &  97.496 & 1 & $798.9\pm0.6$ & 181.6 & $8.29\pm0.07$ & $-13.46\pm0.09$ & 
$-1.95\pm0.05$ & $15.8\pm0.1$ \\
  Gain Calibrator & All &  97.496 & 2 & $799.0\pm0.6$ & 197.9 & $10.50\pm0.10$ & $-12.84\pm0.09$ & 
$-2.01\pm0.05$ & $16.6\pm0.1$ \\
  Gain Calibrator & All &  97.496 & 3 & $799.2\pm0.6$ & 177.0 & $11.70\pm0.10$ & $-12.49\pm0.07$ & 
$-1.80\pm0.05$ & $17.1\pm0.1$
\enddata
 \label{tab:tsplitfluxes} 
 \tablenotetext{a}{The times of the three SB executions (considering target and gain calibrator 
scans only) are: 5.135--5.169, 5.174--5.208, and 5.211--5.245~days, respectively.}
\end{deluxetable*}

\subsection{Polarization measurements}
\label{text:polmeasurements}
We are unable to reproduce the polarization measurements of \cite{uth+19} in our analysis. In 
the lower sideband, our measurements of Stokes $Q$ are statistically indistinguishable from zero, 
whereas Stokes $U$ appears positive, rather than negative as found by the previous authors. 
In the upper sideband, the 30~s amplitude self-calibration yields an extremely large change in 
Stokes $Q$ relative to the 20~min calibration; the Stokes $Q$ flux density changes from 
$-57.2\pm10.6~\mu$Jy to $-7.1\pm10.0~\mu$Jy, highlighting the danger in leaving the cross-hand 
phase 
free while self-calibrating weakly polarized sources. We note that self-calibration moves the $QU$ 
data points closer to the origin in the $Q$-$U$ plane, corresponding to zero polarization 
(Fig.~\ref{fig:QU_GRB}). 

In all cases, our images reveal an unexpected detection in Stokes $V=-69.3\pm10.0~\mu$Jy in the 
LSB and $V=-77.5\pm10.5~\mu$Jy in the USB, corresponding to a circular polarization at the level of 
$\approx0.21\%$--0.24\%. This is similar to the level previously noted for the phase calibrator.
We caution that the minimum systematic uncertainty in circular 
polarization measurements with ALMA is currently $\approx0.6\%$, and hence this (statistically 
significant) detection of Stokes $V$ is almost certainly spurious and most likely indicates 
residual (unremovable) calibration errors. These may arise, for instance, from time-variable $XY$ 
phase or standing waves in the orthomode transducers.
Another possibility is that the polarization
calibrator has non-zero circular polarization. Standard polarization calibration assumes negligible
$V$ in the polarization calibrator. Thus, non-zero $V$ in the calibrator may corrupt the calibration
solution, and the calibrator's $V$ may subsequently appear in calibrated science target data.
We note that $I$ to $V$ conversion due to beam squint is expected to be negligible close to the 
primary beam axis.
The level of spurious circular polarization is similar to that of the claimed 
linear polarization detection in \cite{uth+19}; however, as those authors do not present Stokes $V$ 
images or photometry, we cannot perform a direct comparison in Stokes $V$.  We stress that the 
systematic calibration errors causing a spurious Stokes $V$ signal may or may not be the same 
errors 
causing the spurious $Q$ and $U$ detections we see in the data.

We further test for calibration stability by dividing the data into time bins by each of 
the three executions of the scheduling block.
(Table~\ref{tab:tsplitfluxes}). 
We find that the 
polarization measurements exhibit 
significant time variability. Stokes $Q$ increases by $\approx0.3\%$ of Stokes $I$, changing sign 
during the observation from negative to positive (Fig.~\ref{fig:varQUV_GRB}). The change is 
$\approx10\sigma$ relative to the typical (statistical) measurement uncertainty in $Q$. The 
variation is especially pronounced in the LSB ($\approx0.4\%$ of $I$), and is even larger in the 
LSB for Stokes $Q$ ($\approx0.5\%$ of $I$) prior to self-calibration. At the same time, the 
polarization 
properties exhibit very different structures in the lower and upper sideband. For instance, Stokes 
$U$ is positive in the first scheduling block in the LSB ($89\pm16~\mu$Jy), but negative in the 
USB ($-71\pm14~\mu$Jy). {The difference of $\approx160~\mu$Jy between LSB and USB, a factor 
of $\approx9$ relative to their mean, cannot arise from the $\approx3$\% difference in their Stokes 
$I$.} Our measurements of Stokes $U$ for the GRB decrease toward zero with 
time in both sidebands. This trend is robust to 
self-calibration (Fig.~\ref{fig:QU_GRB}). The time scale of this evolution is $\approx2.6$~hours at 
$\approx5.2$~days. The corresponding fractional duration of only $\approx 2\%$ would imply 
unphysically rapid changes, $\alpha \approx -70$ for an expected power law temporal evolution, 
$P\propto t^{\alpha}$, ruling out intrinsic changes and implying instabilities in the polarization 
calibration. 

We search for systematic calibration errors by repeating the above analysis for the gain 
calibrator, J1130-1449. We self-calibrate the data separately in the two sidebands in the same 
manner as for GRB~171205A. The images reveal statistically significant circular polarization at the 
level $V\approx0.25\%$ (sideband-averaged), similar to that obtained in the GRB data. This source 
has been variously categorized as an optical quasi-stellar object (QSO) and blazar 
\citep{mgl+09,mkl+16}. QSOs and blazars have been observed to circular polarization at the 
$\approx0.1\%$ level at cm wavelengths \citep{rns00}. However, the circular polarization fraction 
is expected to fall with frequency as $V/I\propto \nu^{\alpha_{\rm V}}$ with $-\alpha_{\rm 
V}\approx1$--3 \citep{pac73,mel97}, implying negligible Stokes $V$ at the mm wavelengths employed 
here. Indeed, very few blazars have detected circular polarization at mm wavelengths 
(\citealt{atwk10,atm+18}; however, see also \citealt{tam+18}). 
Thus, the consistent detected Stokes $V$ for the gain calibrator may imply residual uncorrected 
instrumental polarization in the data.

We search the gain calibrator data for systematics by investigating variability in the polarization 
properties in time and frequency. As in the case of the GRB, we find that Stokes $Q$ and $U$ vary 
by up to $0.4\%$ of Stokes $I$ over time, and the variation is as strong as $\approx0.7\%$ of 
Stokes $I$ in the LSB (Fig.~\ref{fig:varQUV_GCal}). Intrinsic variations on the time scale of 
$\approx2.6$~hours as observed here are not expected in radio-loud AGN \citep{den65}. Whereas 
interstellar scintillation can cause variability on much shorter (hour) time scales, this effect is 
expected to be negligible at mm wavelengths \citep{qui92,gn06}. Thus, the observed strong 
variability of the polarization properties of the gain calibrator are most likely instrumental and 
not intrinsic to the source. One possible origin for these systematics may be time-varying
$XY$ phase. However, investigating this requires second-order calibration corrections,
which are beyond the scope of this work.

\subsection{Systematic calibration uncertainty}
In light of the observed variability of the GRB and gain calibrator data in time and frequency, we 
believe the systematic calibration uncertainty for this dataset is larger than the nominal 
$3\sigma$ value of 0.1\% quoted in the ALMA Cycle 4 Technical 
Handbook\footnote{\url{
https://almascience.nrao.edu/documents-and-tools/cycle4/alma-technical-handbook}}, relevant ``for 
the brightest calibrators''. Whereas the Handbook does not clarify this term precisely, calibrators 
with polarization fraction $\gtrsim10\%$ are available to 
ALMA\footnote{\url{http://www.alma.cl/~skameno/AMAPOLA/}}, and thus, with a fractional polarization 
of $\approx6.3\%$, J1256-0547 is only moderately strongly polarized. 

To quantify the true systematic, we use the observed variability in the polarization of the gain 
calibrator, and assume that the calibrator's 
intrinsic polarization is constant with time over our observations. We observe a maximum deviation 
of $\Delta\Pi_{\rm L}\approx0.16$\% of Stokes $I$ for the calibrator when both sidebands are 
combined 
(this number is $\Delta\Pi_{\rm L}\approx0.21\%$ prior to self-calibration). If 
the systematic error is a random (Gaussian) process, then this number would be an 
overestimate of the intrinsic standard deviation of that random process. 
The expectation value of 
the difference between the maximum and minimum 
(i.e., the range\footnote{This quantity follows a Gumbel distribution.})
of three numbers drawn from a unit normal distribution is 
$3\pi^{-1/2}\approx1.69$ \citep{sch06a}.
Thus, we estimate an additional $1\sigma$ systematic calibration uncertainty of
$\approx0.16\%/1.69\approx0.09\%$ for these observations. 

In conjunction with the statistical uncertainty of $\approx0.033\%$ in the linear polarization 
measurement of the GRB when all the data are combined (Table~\ref{tab:polcal}), the total 
($1\sigma$) uncertainty in the polarization measurement is $\approx0.10\%$. This yields a linear 
polarization measurement of $41.7\pm32.4~\mu$Jy (undebiased), and thus the detection of 
polarization in this event is only significant at $\approx1.3\sigma$. Since $P/\sigma_{\rm 
P}\lesssim\sqrt{2}$, the maximum likelihood estimate for $P$ is $\hat{P}=0$ upon correcting for 
Rician bias \citep{vai06}.  
Even for the total linearly polarized density of $\approx87~\mu$Jy reported in \cite{uth+19}, the 
addition of a $0.09\%$ systematic uncertainty renders the measurement at best a $\approx2.7\sigma$ 
detection. Given the significant variability observed and our inability to reproduce the earlier 
authors' results using an independent analysis, we consider these data to provide a $3\sigma$ 
upper limit of $\lesssim0.30\%$ (combining systematic and statistical uncertainty, 
and corresponding to $P\lesssim97.2~\mu$Jy) 
on the linear polarization of GRB~171205A for the remainder of this work.

\begin{figure}
  \centering 
   \includegraphics[width=0.5\textwidth]{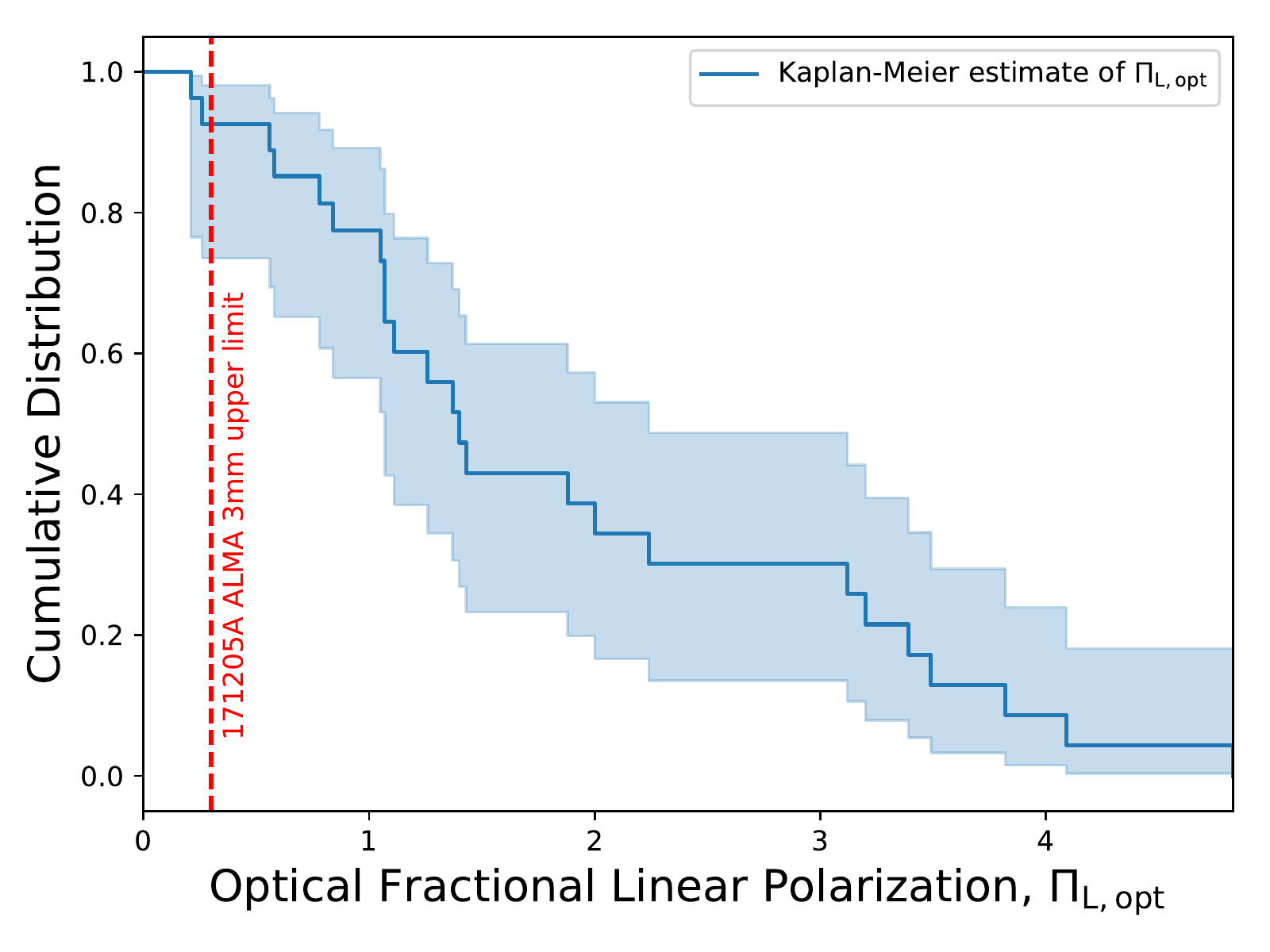}
  \caption{Kaplan-Meier cumulative distribution function of the optical linear polarization of GRB 
afterglows observed between 2.6 and 10.4 days (i.e., within a factor of 2 in time relative to the 
ALMA observations of GRB~171205A), including polarization upper limits, from \cite{cg16}. Between 
4--27\% of optical afterglow polarization measurements are lower than the ALMA 3mm upper limit of 
$<0.30\%$ for GRB~171205A (intersection of the dashed line with the shaded region). Thus, we cannot 
rule out that the linear polarization in this burst is intrinsically low.}
\label{fig:optpol}
\end{figure}

\begin{figure*}
 \centering
 \begin{tabular}{cc}
  \includegraphics[width=\columnwidth]{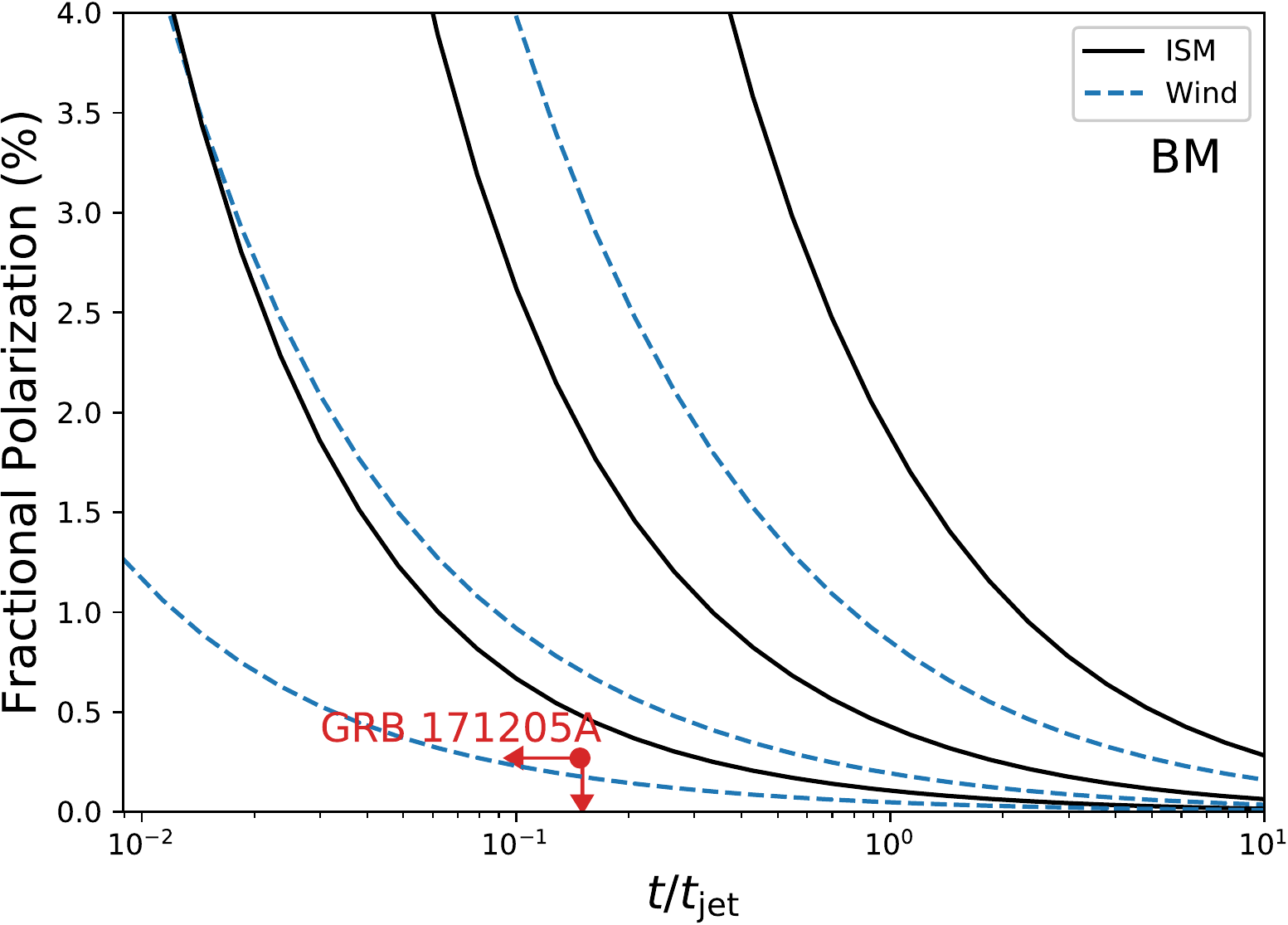} &
  \includegraphics[width=\columnwidth]{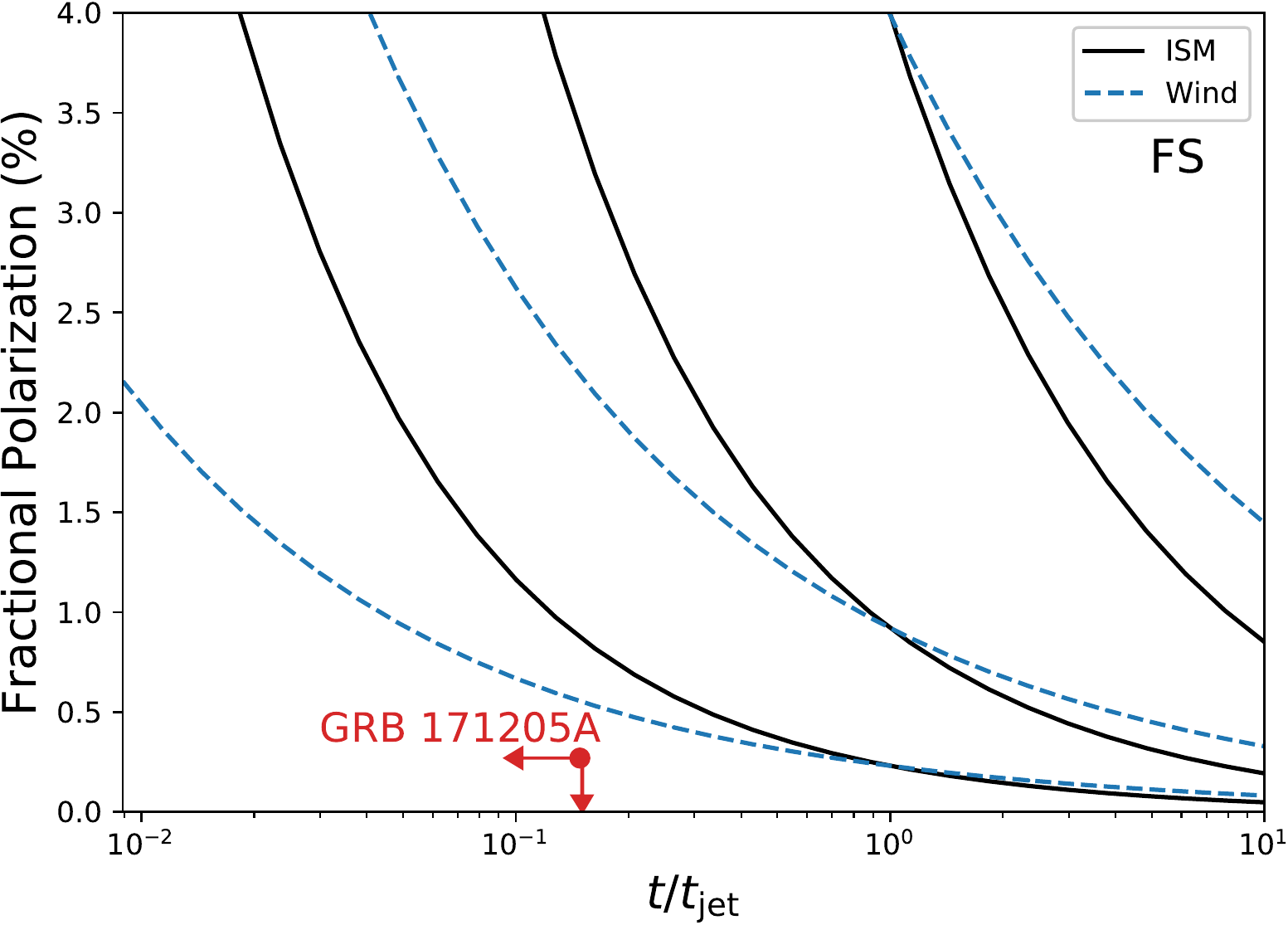} \\
 \end{tabular}
 \caption{The expected polarization signature from uniform jets with toroidal magnetic fields 
expanding into constant density (solid lines) and wind-like environments (dashed lines) for a 
Blandford-McKee (BM) evolution (left) and FS-like evolution (right), together with the measured 
polarization fraction upper limit for GRB~171205A (red point). The lower limit on $\tjet$ 
corresponds to $t/\tjet\lesssim0.15$ for the GRB. The three lines are for $\theta_{\rm 
obs}/\thetajet=0.2$, 0.1, and 0.05, from highest to lowest, respectively. A toroidal magnetic 
field geometry is difficult to reconcile with the upper limit for most viewing angles if the 
emission arises from a RS.}
\label{fig:poltheory}
\end{figure*}

\section{Discussion}
\label{text:results}
The precise interpretation of the polarization upper limit depends strongly upon whether the 
emission arises 
from shocked jet material (i.e., the reverse shock; RS) or from the shocked ambient environment 
(the 
forward shock; FS), and upon the magnetic field structure in the region of emission 
\citep{gl99,gk03,rlsg04,gt05}. 
A detailed study of the afterglow emission 
and its decomposition into forward and reverse shock components is beyond the scope of this work, 
but we briefly discuss both scenarios. 
In the case of radiation powered by FS emission and where polarization is the result of
viewing a region with shock-produced magnetic fields off-axis, the temporal evolution 
of the polarization fraction typically exhibits two peaks;  however, the polarization fraction can 
be very low, especially when the viewing geometry is close to being on-axis \citep{rlsg04}. Thus, 
we cannot rule this scenario out. 

\subsection{No strong evidence for thermal electrons}
A suppression of the polarization by Faraday depolarization due 
to a quasi-thermal population of electrons not accelerated at the FS, as argued by 
\cite{uth+19}, {is} an interesting possibility \citep{tin08}. 
{In their analysis of this burst, \cite{uth+19} contrast their reported ALMA Band 3 
measurement of $\Pi_{\rm L}=(0.27\pm0.04)\%$ with optical polarization observations \citep{cg16}. 
They claim that optical observations during the FS-dominated phase yield a weighted average optical 
linear polarization of $\Pi_{\rm L}\approx1.2$\% (without error bars; they also do not describe how 
they remove any potential RS contamination). They ascribe the difference between the measured and 
the ``typical'' optical polarization to the presence of quasi-thermal electrons.}

{Whereas such a population should indeed exist \citep{ew05,ss11}, we caution that (i) there 
is no evidence that radio polarization measurements track the optical polarization (indeed, there 
is 
exactly one radio polarization detection of a GRB afterglow to date, with $\Pi_{\rm 
L,opt}\approx3\Pi_{\rm L,radio}$; however, the detected optical polarization for that event is 
likely dominated by extrinsic dust scattering; \citealt{lag+19,jmk+20}); and (ii) the \cite{uth+19} 
analysis ignores the optical polarization upper limits. Including these upper limits, we find that 
as many as 27\% of optical polarization observations made within a factor of 2 in time of the time 
of these ALMA observations (5.19~days, corresponding to the range $\approx2.6$--10.4~days) are 
below the ALMA 3\,mm polarization upper limit (Fig.~\ref{fig:optpol}). Thus, it is entirely 
possible that the optical polarization in this burst may have been intrinsically lower than the 
observed radio upper limit.} 
Futhermore, we note that polarization levels approaching zero can be expected from purely 
shock-generated fields. {Thus,} the data do not provide direct observational evidence for 
non-accelerated particles.

\subsection{Constraints on magnetic field geometry}
We now discuss the observed upper limit on the polarization at $\approx5.19$~days in the context of 
the magnetic field geometry in the jet powering GRB~171205A. 
In general, the observed polarization degree is a function of the ratio of the off-axis viewing 
angle ($\theta$) to the opening angle of the jet ($\thetajet$), 
and the time relative to the jet break time, $\tjet$ 
\citep{rho99,sph99}. The X-ray light curve for the afterglow of GRB~171205A exhibits a shallow, 
unbroken power law decay with $\alpha\approx-1.06$ to 
$\gtrsim35$~days\footnote{\url{https://www.swift.ac.uk/xrt_live_cat/00794972/}}, indicating that 
$\tjet\gtrsim35$~days. Thus, we consider the observation time of $t_{\rm obs}=5.19$~days to 
correspond to an upper limit on the ratio $t_{\rm obs}/\tjet \lesssim0.15$~days. 

Together with coeval Atacama Compact Array (ACA) 345~GHz observations, the ALMA 97.5~GHz data 
indicate an optically thin spectrum in the mm-band at $\approx5.19$~days, for which we calculate 
$\beta_{\rm mm}=-0.51\pm0.04$. {On the other hand, the spectral index between the LSB 
and USB within Band 3 is lower, $\beta_{\rm 3mm}=-0.25\pm0.01$, indicating that a spectral break 
frequency lies not too far below ALMA Band 3. VLA observations at 5--16~GHz around the same time 
($\approx4.3~$days) exhibit a steeply rising spectrum, with $\beta_{\rm cm}=1.46\pm0.03$ 
\citep{uth+19}. These observations indicate that both the synchrotron peak frequency ($\numax$) and 
self-absorption break ($\nua$) are at a frequency lower than ALMA Band 3. Furthermore, the 
VLA spectrum is shallower than the fully self-absorbed expectation of $2\le\beta_{\rm cm}\le2.5$, 
implying that $\nua$ is in the cm band at $\approx5$~days, and that the potential spectral peak 
near the ALMA band is due to $\numax$. Therefore depolarization due to synchrotron 
self-absorption in the ALMA bands is unlikely, indicating that the polarization of the observed 
radiation is intrinsically low.}

{We note that RS emission has been seen in ALMA observations of GRB afterglows as late as 
$\approx4$~days after the burst} \citep{lab+18,lves+19}.
{If} the {radio} emission {in GRB~171205A} arises from adiabatically 
cooling, reverse-shocked ejecta, then the polarization upper limit presents strong constraints on 
the magnetic field structure in the GRB jet. 
For a magnetic field ordered on patches of scale, $\theta_{\rm B}$, the observed polarization would 
be suppressed by a factor of the number of patches visible, $N\approx(\Gamma\theta_{\rm B})^{-2}$, 
where $\Gamma$ is the jet Lorentz factor at the time of observations \citep{no04,gt05}. This 
implies 
$\theta_{\rm B}\lesssim \Pi_{\rm L,lim}\Gamma^{-1}\Pi_{\rm 
max}^{-1}\approx4\times10^{-3}\Gamma^{-1}$~rad, where we have taken $\Pi_{\rm 
L,max}=(1-\beta)/(5/3-\beta)\approx0.68$ \citep{gt05} for $\beta\approx-0.43$ \citep{uth+19}. This 
limit is consistent with the value of $\theta_{\rm B}\approx10^{-3}$~rad inferred from polarization 
observations of the reverse shock in GRB~190114C \citep{lag+19}. Thus, if the emission arises from 
the reverse shock, this may indicate a universal magnetic field coherence scale. 

The low degree of polarization disfavors models of polarization produced by toroidal magnetic 
fields in GRB jets. To see this, we compare the models of \cite{gt05} together with the data in 
Fig.~\ref{fig:poltheory}. We explore a range of off-axis angles and both constant density and 
wind-like progenitor environments. For the Lorentz factor evolution of the ejecta after 
deceleration ($\Gamma\propto R^{-g}$), we consider two scenarios: a minimum value of $g=(3-k)/2$, 
corresponding to the evolution of the fluid just behind the forward shock, and $g=7/2-k$, a maximum 
value expected for a reverse shock, corresponding to the Blandford-McKee self-similar solution 
\citep{kob00,gt05}\footnote{Here $k$ is the power law index of the radial density profile.}. In the 
former case, we find that a toroidal field would produce too high a polarization degree regardless 
of viewing angle or the circumburst geometry. In the latter case, $\theta_{\rm 
obs}/\thetajet\approx0.05$ is marginally allowed by the data; however, this would require a very 
precise alignment of the jet axis with the line of sight, and is therefore unlikely\footnote{This 
would require a chance alignment probability of $\approx2\times10^{-5}$ for a typical opening angle 
of $\approx10^\circ$.}. Finally, we note that a ``universal structured jet'' model 
with a toroidal magnetic field can be ruled out, since it would produce a much higher degree of 
polarization, $\Pi_{\rm L}\gtrsim30\%$ at $t\lesssim\tjet$ \citep{lcg+04}.

\begin{figure*}
 \centering
 \begin{tabular}{ccc}
  \includegraphics[width=0.31\textwidth]{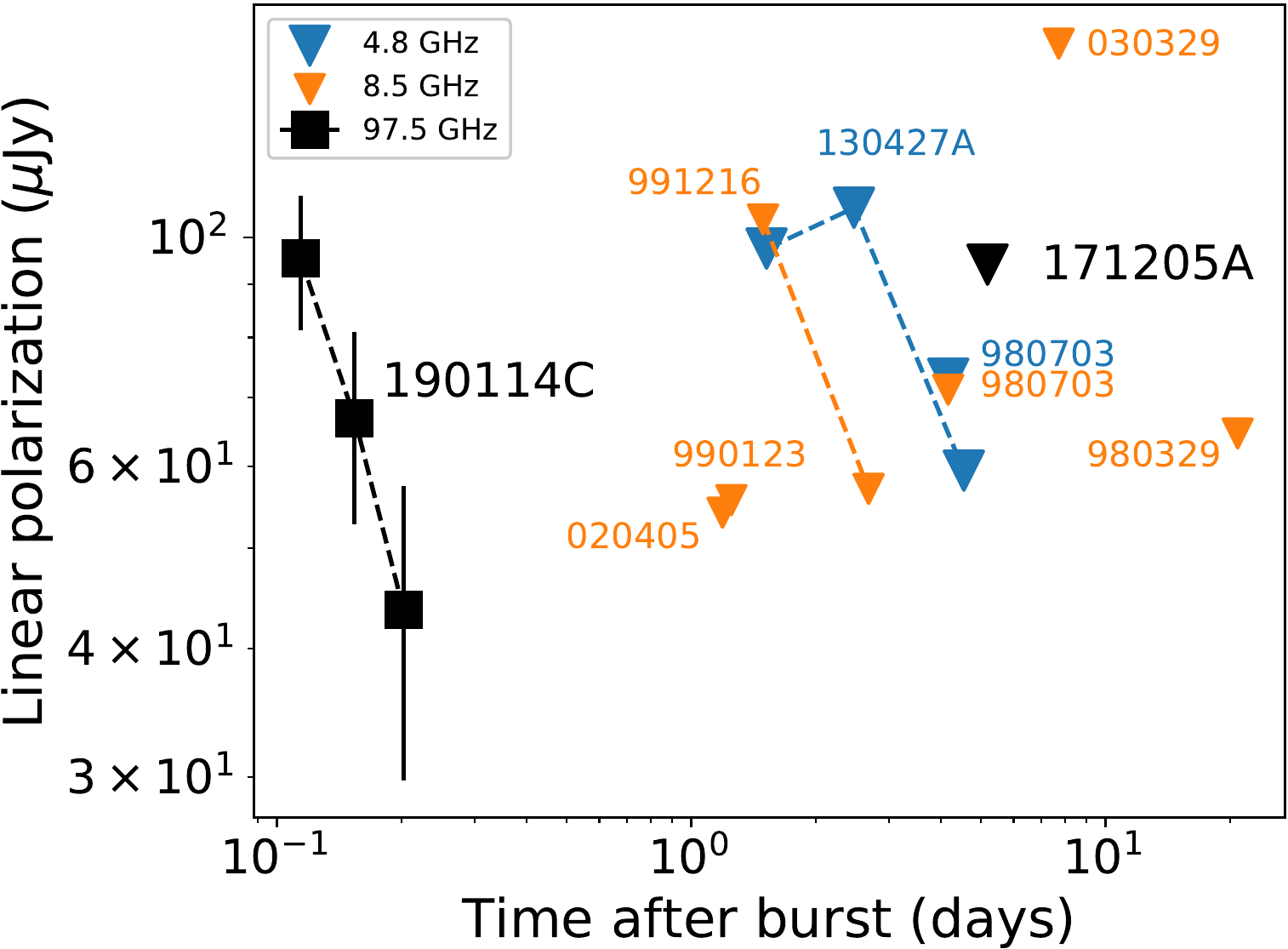} &
  \includegraphics[width=0.31\textwidth]{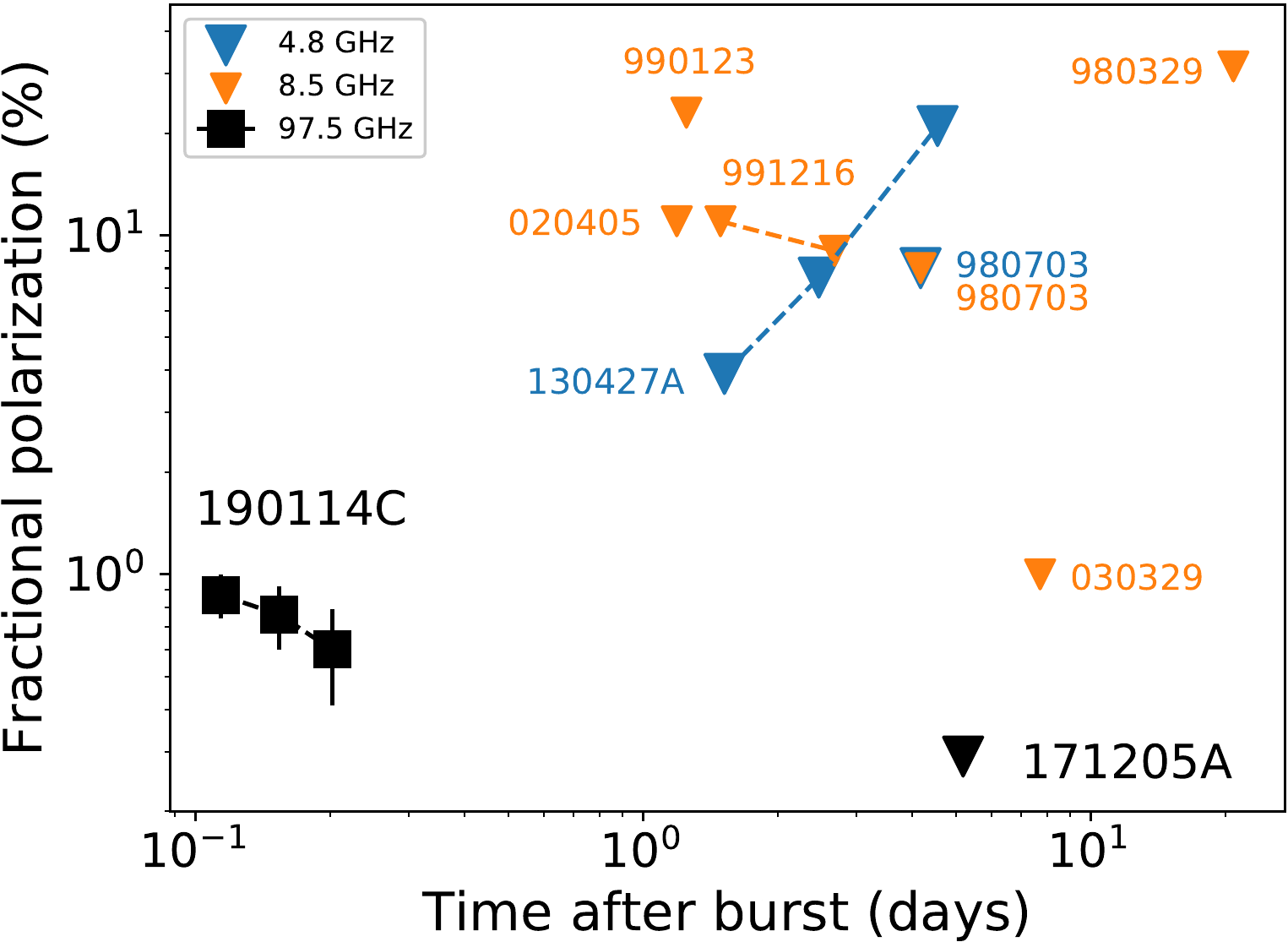} & 
  \includegraphics[width=0.31\textwidth]{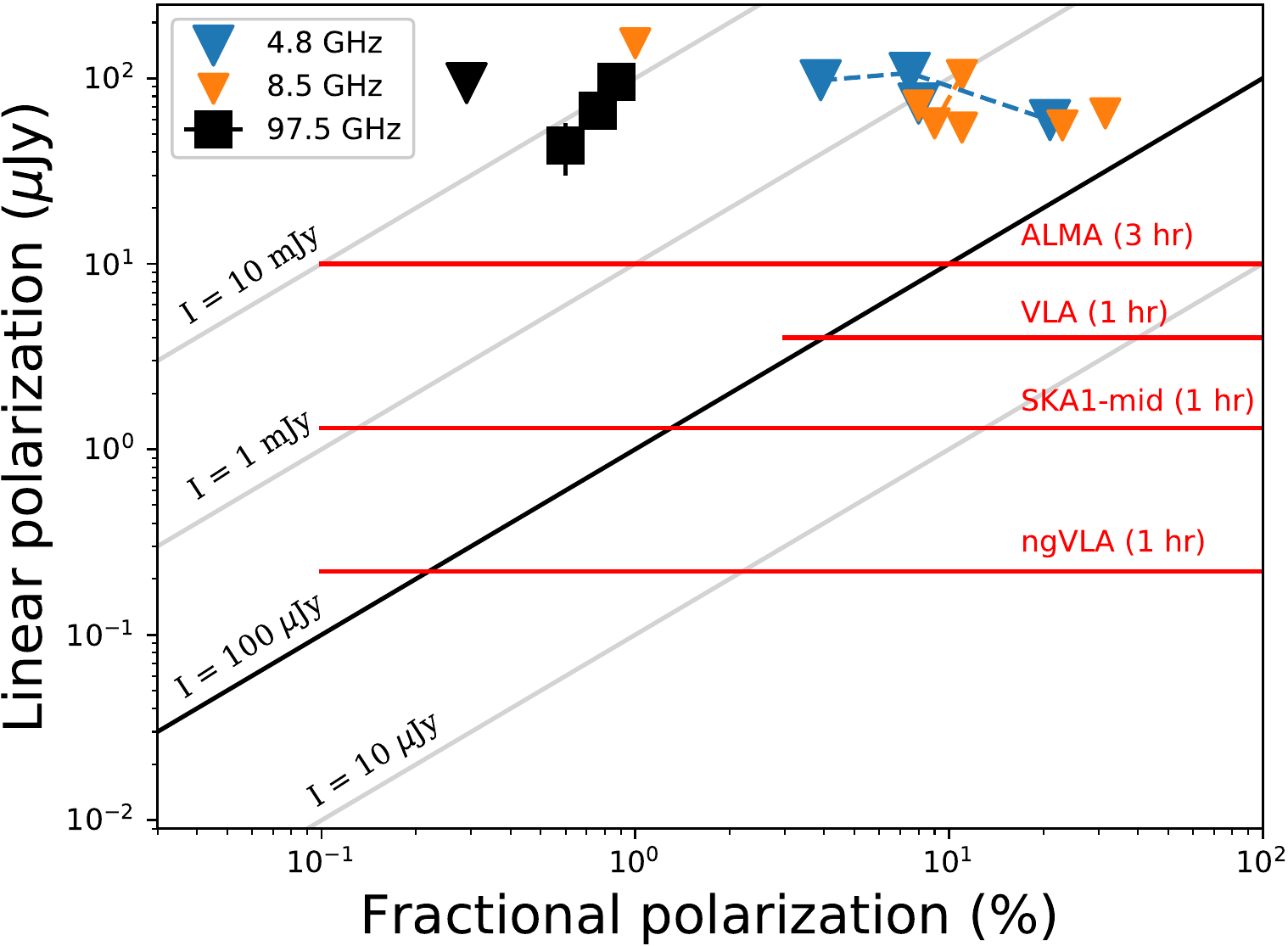} \\
 \end{tabular}
 \caption{Linear polarized intensity (left) and fractional linear polarization (center) as a 
function of observation time for GRB radio afterglows at $\approx4.8$~GHz (blue), 
$\approx8.5$~GHz (orange) and $\approx97.5$~GHz (black) from this work and collected from the 
literature \citep{tfk+98,fyb+03,tfbk04,gt05,vdhpdb+14,lag+19}. The mm-band observations are from 
ALMA, while the cm-band upper limits are from the VLA, European VLBI Network (130427A), and the 
Very 
Long Baseline Array (030329). The bright afterglows of GRB~030329 and 171205A allow for strong 
upper limits on the fractional polarization, even though the individual upper limits on the 
polarized intensity are similar to those of other bursts. However, the late observations in these 
cases most likely precluded a detection. We plot the linear polarized intensity as a function of 
the polarization fraction in the right panel to compare with the sensitivity of the current and 
upcoming generations of facilities for a range of Stokes $I$ values from 10~$\mu$Jy to 10~mJy. For 
typical GRB radio afterglows ($I\approx100~\mu$Jy), polarization observations will be possible with 
SKA1-mid at the level of $\Pi_{\rm L}\approx1\%$, but will require full SKA or ngVLA sensitivity 
for polarization detections at the $\lesssim0.1\%$ level.}
\label{fig:historicpol}
\end{figure*}

\subsection{Radio Polarization of GRB Afterglows}
We now compare this derived upper limit to values previously reported for radio observations of GRB 
afterglows. Our compiled sample of radio linear polarization observations includes one detection 
\citep[GRB~190114C;][]{lag+19}, and several upper limits (GRB~980329, \citealt{tfk+98}; GRB~980703, 
\citealt{fyb+03}; GRBs~990123, 991216, and 020405, \citealt{gt05}; GRB~030329, \citealt{tfbk04}; 
and 
GRB~130427A, \citealt{vdhpdb+14}; and GRB~171205A, this work). \
We convert upper limits listed at different confidence intervals 
to a uniform $3\sigma$ limit for comparison across events. We multiply the quoted fractional 
polarization upper limits by the Stokes $I$ flux density to estimate the upper limit in flux 
density 
units, and plot these separately at C-band ($\approx4.8$~GHz), X-band ($\approx8.5$~GHz), and at 
3~mm ($\approx97.5$~GHz) in Figure~\ref{fig:historicpol}.

GRB~171205A exhibited the brightest Stokes $I$ flux density of our sample at the time of the 
polarization observations. Thus, our upper limit on the polarized flux of GRB~171205A, while not 
the strongest in absolute flux terms, yields the deepest upper limit on the fractional polarization 
of $\Pi_{\rm L}<0.30\%$ (including systematics). This imposes a factor of 3 stronger 
constraint on the intrinsic polarization of radio afterglows than previously performed 
\citep[GRB~030329;][]{tfbk04}. As discussed in \cite{uth+19}, the emission appears to be optically 
thin at 97.5~GHz at 5.19~days. Therefore, depolarization due to synchrotron self-absorption is 
unlikely to be the cause for the non-detection of polarized emission \citep{tin08,gvdh14}, 
suggesting that the absence of strongly polarized emission is intrinsic to the source.

We note that the polarization upper limits (and the measurement in the case of GRB~190114C) all 
lie in the range of $\approx40$--$150~\mu$Jy; the difference in the polarization fractions arises 
from the large spread (over two orders of magnitude) in Stokes $I$ flux densities in the respective 
bands at the time of observation. Of these, the ALMA observations of GRB~190114C represent the 
earliest post-burst polarization-sensitive observations obtained for any mm-band GRB afterglow. The 
fractional polarization limits in the cm-band are all higher (i.e., worse) than those obtained in 
the mm-band, indicating the need to improve instrument sensitivity and stability at these 
frequencies 
in order to probe polarized emission from GRB afterglows.

According to the ngVLA reference design, the $1\sigma$ point source sensitivity at 8~GHz in 1~hour 
of on-source integration is $\approx0.22~\mu$Jy. The polarization sensitivity in the current design 
is expected to be better than $\Pi_{\rm L}\approx0.1\%$. The required on-source time to detect a 
typical $\approx 100~\mu$Jy GRB afterglow \citep{cf12} polarized at $0.1\%$ will be 
$\approx5$~hours, although the source polarization may vary over this same period \citep{lag+19}. 
The full Square Kilometer Array (SKA2) would achieve a similar sensitivity; however, in phase 1, 
the SKA-mid would require upwards of $56$~hours for a typical GRB radio afterglow.

A detection of linearly polarized radio emission unambiguously associated with the afterglow 
forward shock would provide the first constraints on the magnetic field structure and viewing 
geometry for long-duration GRBs \citep{gk03}. In particular, the evolution of this 
quantity across the jet break is a sensitive measure of the degree of order in the magnetic fields, 
the jet structure, and the off-axis viewing angle \citep{rlsg04}. Thus, we suggest that a more 
robust interpretation of afterglow polarization requires sensitive measurements (with detections) 
at multiple epochs. Such observations, while challenging for typical GRBs with ALMA in the mm band, 
may be routinely tractable with the ngVLA and full SKA.

\section{Conclusions}
We have presented a series of tests useful for estimating the impact of systematic calibration 
errors in ALMA polarization data. {In particular, we recommend basic sanity checks of (i) 
dividing the data in time and frequency to test for calibration stability, and (ii) checking the 
gain calibrator or using test calibrators, if available, to verify and quantify the success of 
polarization calibration.} While these tests have been performed at 3~mm here, they are widely 
applicable to {observations} at any frequency.

We have re-analyzed ALMA Band 3 (3~mm) full continuum polarization observations of GRB~171205A 
taken $\approx5.19$~days after the burst {and performed detailed verification steps to test 
the stability of polarization calibration.} In contrast to {previous work} \citep{uth+19}, 
we do not detect significant linear polarization from the radio afterglow. 
We find a higher systematic uncertainty than assumed by \cite{uth+19}, and infer a $3\sigma$ upper 
limit of $P\lesssim97.2~\mu$Jy, corresponding to $\Pi_{\rm L}\lesssim0.30$\% of Stokes $I$, for 
which we derive a value of $32.44\pm0.03$~mJy (statistical error). 
{The upper limit on $\Pi_{\rm L}$ is consistent with the range of optical linear 
polarization observed for GRB afterglows, and thus not immediately indicative of the presence of a 
population of thermal electrons}. 
If the emission arises in the reverse-shocked region, the 
upper limit rules out a toroidal magnetic field geometry for most viewing angles, and is consistent 
with random magnetic field patches of coherence length, $\theta_{\rm B}\lesssim4\times10^{-3}$~rad. 
We have compiled observations of polarized emission in GRB radio afterglows from the literature, 
and 
demonstrate that the current observations and limits of linear polarized intensity span a narrow 
range, likely due to signal-to-noise limitations. 
We expect that improvements in cm-band polarization sensitivity and stability, such as with the 
ngVLA and full SKA, will open a new avenue for pursuit of GRB jet structure and 
magnetization in the future. 

\acknowledgements
{We thank the anonymous referee for their suggestions, which improved this manuscript.}
TL thanks R.~Margutti and P.~Schady for helpful discussions. 
CLHH acknowledges the support of both the NAOJ Fellowship as well as JSPS KAKENHI grant 18K13586.
This paper makes use of the following ALMA data: ADS/JAO.ALMA\#2017.1.00801.T. ALMA is a 
partnership of ESO (representing its member states), NSF (USA) and NINS (Japan), together with NRC 
(Canada), MOST and ASIAA (Taiwan), and KASI (Republic of Korea), in cooperation with the Republic of 
Chile. The Joint ALMA Observatory is operated by ESO, AUI/NRAO and NAOJ.
The National Radio Astronomy Observatory is a facility of the National Science Foundation 
operated under cooperative agreement by Associated Universities, Inc.


\bibliographystyle{apj}

\end{document}